\documentclass[onecolumn,aps,prl,preprint,groupedaddress,showpacs]{revtex4-1}
\usepackage{dcolumn}
\usepackage{bm}
\usepackage{amsmath}
\usepackage{amsfonts}
\usepackage{graphics}
\usepackage{graphicx}
\usepackage{color}

\begin{document}


\title{Correlation, Network and Multifractal Analysis of Global Financial Indices}



\author{Sunil Kumar$^{1,2}$}
\author{Nivedita Deo$^{1}$}
\email{ndeo@physics.du.ac.in (corresponding author)}

\affiliation{$^1$Department of Physics $\&$ Astrophysics, University of Delhi, Delhi-110007, India\\
             $^2$Department of Physics, Ramjas College, University of Delhi, Delhi-110007, India}


\date{\today}

\begin{abstract}
We apply RMT, Network and MF-DFA methods to investigate correlation, network and multifractal properties of 20 global financial
indices. We compare results before and during the financial
crisis of 2008 respectively. We find that the network method gives more useful information about the formation of clusters as
compared to results obtained from eigenvectors corresponding to second largest eigenvalue and these sectors are formed on the
basis of geographical location of indices. At threshold 0.6, indices corresponding to Americas, Europe and Asia/Pacific disconnect and form different clusters before the crisis but during the crisis, indices corresponding to Americas and Europe are combined together to form a cluster while the Asia/Pacific indices forms another cluster. By further increasing the value of threshold to 0.9, European countries France, Germany and UK constitute the most tightly linked markets. We study multifractal properties of global financial indices and find that financial indices corresponding to Americas and Europe almost lie in the same range of degree of multifractality as compared to other indices. India, South Korea, Hong Kong are found to be near the degree of multifractality of indices corresponding to Americas and Europe. A large variation in the degree of multifractality in Egypt, Indonesia, Malaysia, Taiwan and Singapore may be a reason that when we increase the threshold in financial network these countries first start getting disconnected at low threshold from the correlation network of financial indices. We fit Binomial Multifractal Model (BMFM) to these financial markets.
\end{abstract}


\maketitle

\section{Introduction}
Over the last few years, there has been a growing interest of physicists in economic systems \cite{ecophy00,sengupta99}. The
Random Matrix Theory (RMT) was developed \cite{guhr98,mehta91,bowick91} to deal with the statistics of eigenvalues and
eigenvectors of complex many-body systems and it has been successfully used to investigate phenomena from different areas
such as quantum field theory, quantum chaos, disordered systems etc. and recently to a large number of financial
markets \cite{bouchaud01,gopistanely99,mantegnanature95,laloux99,plerou99,pleroupre99,plerou02,gobaix03,gopihes01,kulkarni07,
sitabhra07,marsili98,shen09,wang11,meric08,junior12,conlon09} to investigate the structure of cross-correlations in financial
markets. The few largest eigenvalues deviate significantly from the RMT prediction. The largest eigenvalue represents the
collective information about the correlation between different stocks and its trend is expected to be dependent on the market
conditions, whereas the component of eigenvectors corresponding to remaining large eigenvalues are associated with the formation
of different sectors in financial market. Complex network technique in nature have become important method for studying properties of complex systems in the real world and penetrated into statistical physics, social sciences, biological sciences,
financial markets \cite{barabasi02, estrada10, song11,newman99,watts98,barrat00,erdos1960,huang2009} and many other fields. The
study of complex networks has been initiated by a desire to understand various real systems from the empirical
data \cite{barabasi02}. Complex network display the spatial topological structure of a system, while the time series is the
expression of the temporal dynamics. As one of most important advances in statistical physics, complex network theory has become
a powerful tool for analyzing financial time series. In this paper, we use threshold and hierarchical method to construct the
correlation network of financial indices. The network generated by threshold method \cite{huang2009} are in general disconnected.
If the system present a clear cluster organization, threshold methods are typically able to detect them. One of the most common
algorithms to detect a possible hierarchical structure hidden in the data is given by the Minimum Spanning
Tree (MST) \cite{mantegna99,mantegna00} and has been applied successfully \cite{bonanno03,onnela03a,onnela03b,coelho07}. This
method selects only
the indices with closest interactions among all indices and it generates a visual presentation of the linkage
relationship among selected interactions between financial indices \cite{brida08,eom07,eom09,eom10,cukur07,jung06,jung08,
tumminello07,corronnello05}. The MST performs better role in
identifying the economic sectors from the correlation matrix when it is compared with other more traditional methods, such as
spectral methods \cite{corronnello05}. In the later procedure one extract the eigenvectors of the correlation matrix and
identifies sectors as groups of indices which have a large component (compared to others) in an eigenvector. Despite
the fact that this method gives some useful information \cite{gopihes01}, the eigenvectors sometimes mix different
economic sectors (especially when eigenvalues are close to one another).
To find multifractal properties in a given financial time series, the Multi-Fractal Detrended Fluctuation
Analysis (MF-DFA) \cite{kantelhardt02} is a robust and powerful technique which identifies and quantifies the multiple
scaling exponents within a time series. It has been successfully applied in different and heterogeneous scientific fields
to study the multifractal properties \cite{peng94,kantelhardt01,ivanov,stanleynature98,kumardeo09,marsili07,sdrozdz05}. To
find the origin of multifractality in time series one can compare the MF-DFA results for the original time series with those
of shuffled and surrogated time series \cite{kantelhardt02,wxz09,zunino08,zunino09}. There are two types of multifractality in
the series: (i) multifractality due to a broad probability density function for the values of the time series, this type of
multifractality cannot be removed by shuffling the series (ii) multifractality due to different long range correlations for
small and large fluctuations, here the corresponding shuffled series will exhibit non-multifractal scaling, since all
long-range correlations are destroyed by the shuffling procedure. If both kind of multifractality are present in a given
series than the shuffled series will show weaker multifractality than the original one.

In this paper, we apply the RMT, complex network and MF-DFA method to the global financial indices
and study financial indices before and during the global financial crisis of 2008 \cite{stanley10,bpstanley10} and second, we
study multifractal properties of $20$ financial indices.
This paper is organized as follows: In Section II, we discuss the data. Section III describes
the RMT approach and results. In Section IV, we discuss the construction and analysis of complex network of financial indices
using threshold and MST method. The MF-DFA method to study multifractal properties and its application to global financial indices are discussed in Section V. Finally we conclude in Section VI.

\section{Data Analyzed}
We analyze the daily closing prices of 20 financial markets around the world traded from the period July 2, 1997 to June 1, 2009, resulting in 3088 returns. These indices are as follows: Argentina: MERV, Brazil: BVSP, Egypt: CCSI, India: BSESN, Indonesia: JKSE, Malaysia: KLSE, Mexico: MXX, South Korea: KS11, taiwan: TWII, Australia: AORD, Austria: ATX, France: FCHI, Germany: GDAXI,
Hong Kong: HSI, Israel: TA100, Japan: N225, Singapore: STI, Switzerland: SSMI, UK: FTSE, and US: GSPC. The data has been
obtained from \cite{yahoo}. There are differences in public holidays or weekends among countries so we shifted the data
according to the rule that when more than $30\%$ of markets did not open on a particular/certain day, we remove that day
from the data, and when it is below $30\%$, we kept existing indices and inserted the last closing price for
each of the remaining indices. Also these markets do not operate at the same time zones. It has been
studied \cite{meric08, junior12, sitabhra07} that correlations of Asian with the USA indices increases when one considers
the correlation of the USA indices with the next day indices of the Asian market. We did not considered weekly data to
avoid the problem of different operating hours between international market so that we do not miss major changes in markets
which tend to occur during a small interval of days. Thus, we consider all indices taken at the same date and filtered the data
accordingly.
\section{Random Matrix Theory Approach}
\begin{figure}[h]
\centering
\includegraphics[scale=0.4]{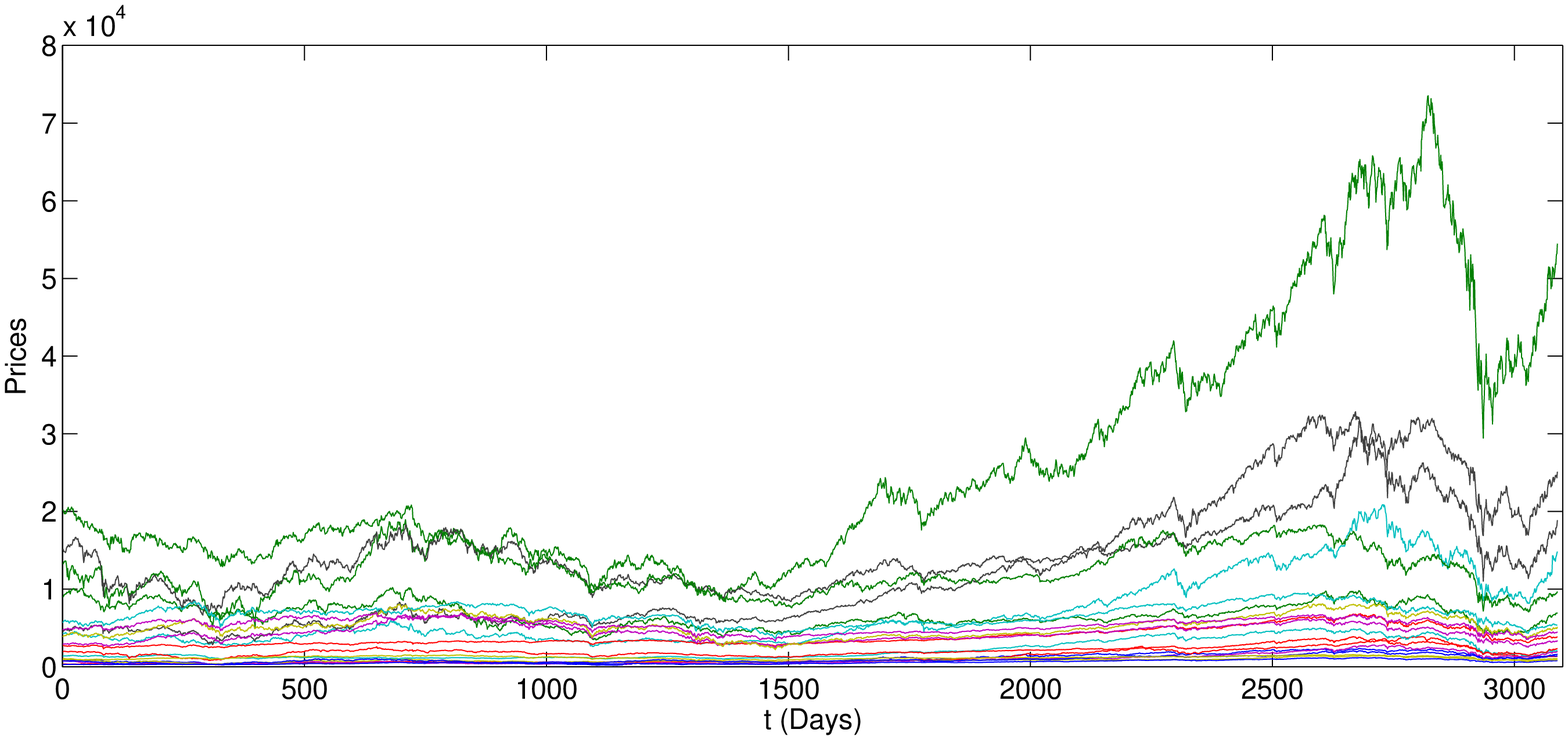}
\includegraphics[scale=0.4]{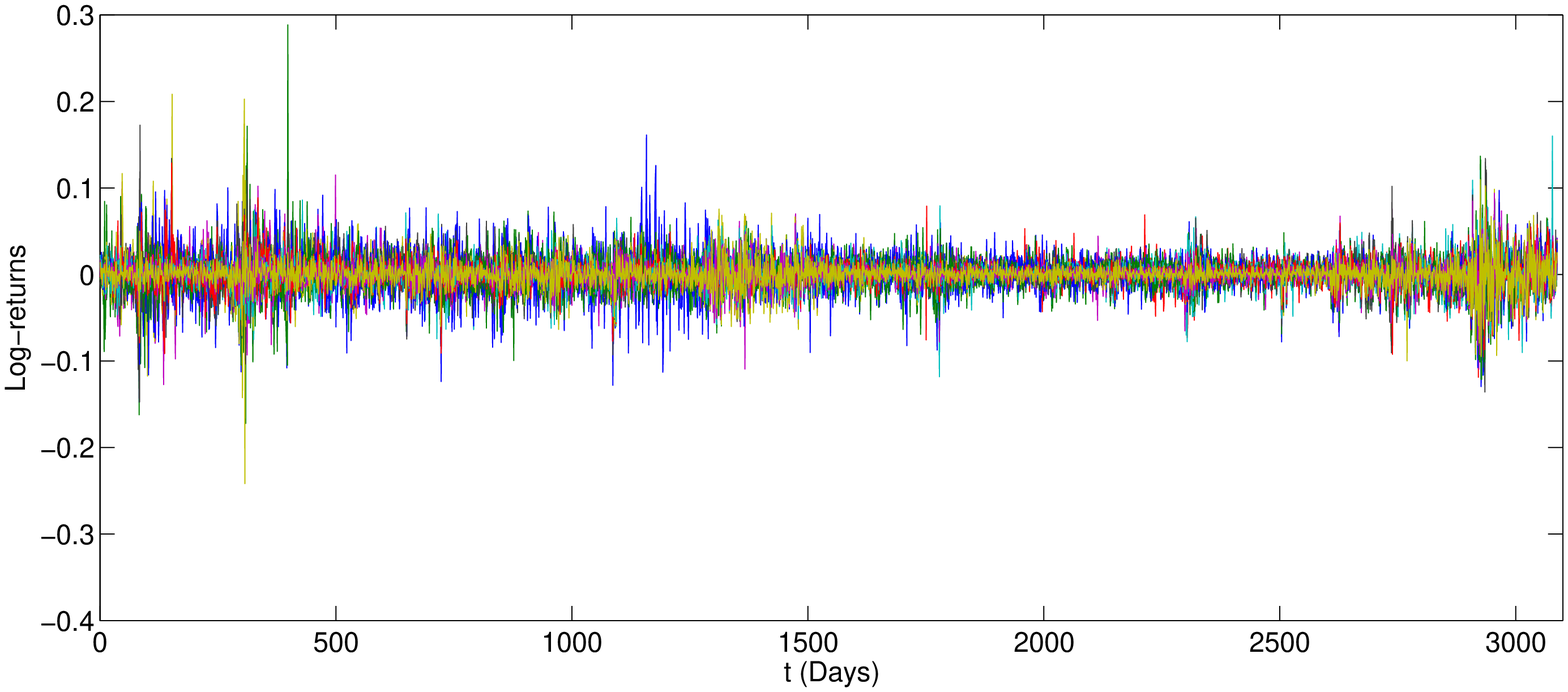}
\caption{(a) Daily closing prices of financial indices of 20 countries
for the period July, 1997 to June, 2009 (b) Corresponding log-returns.}
\label{pricesret}
\end{figure}

\begin{figure}
\centering
\includegraphics[scale=0.4]{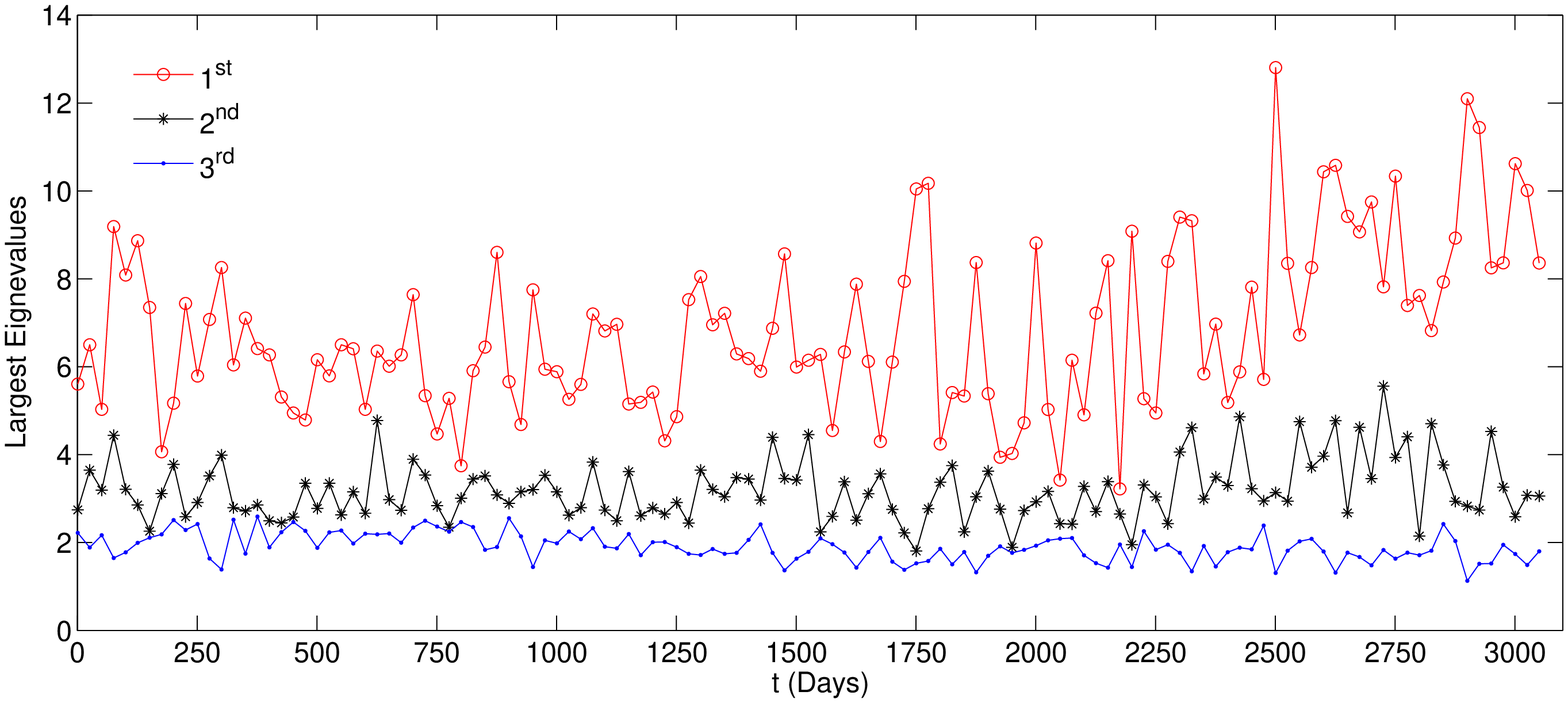}
\caption{Largest eigenvalues of the correlation matrices constructed from daily returns of 20 financial indices
using a sliding window of 25 days.}
\label{largestev}
\end{figure}

\begin{figure}
\centering
\includegraphics[scale=0.4]{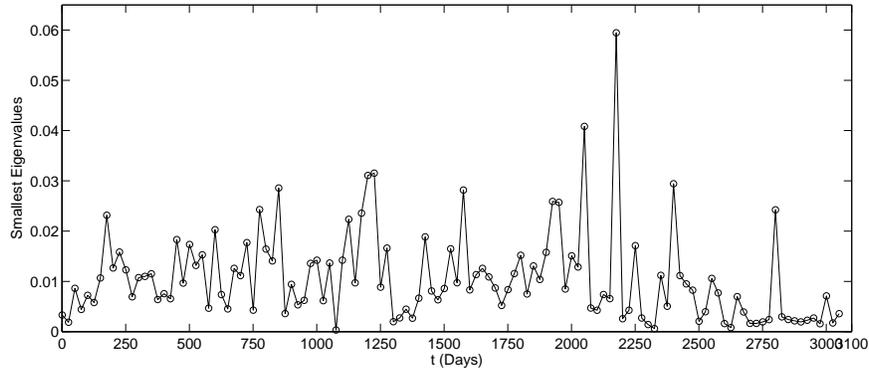}
\caption{The dynamics of smallest eigenvalues of the cross-correlation matrices constructed
from 25 days time window for all financial indices.}
\label{smallestev}
\end{figure}

\begin{figure}
\centering
\includegraphics[scale=0.4]{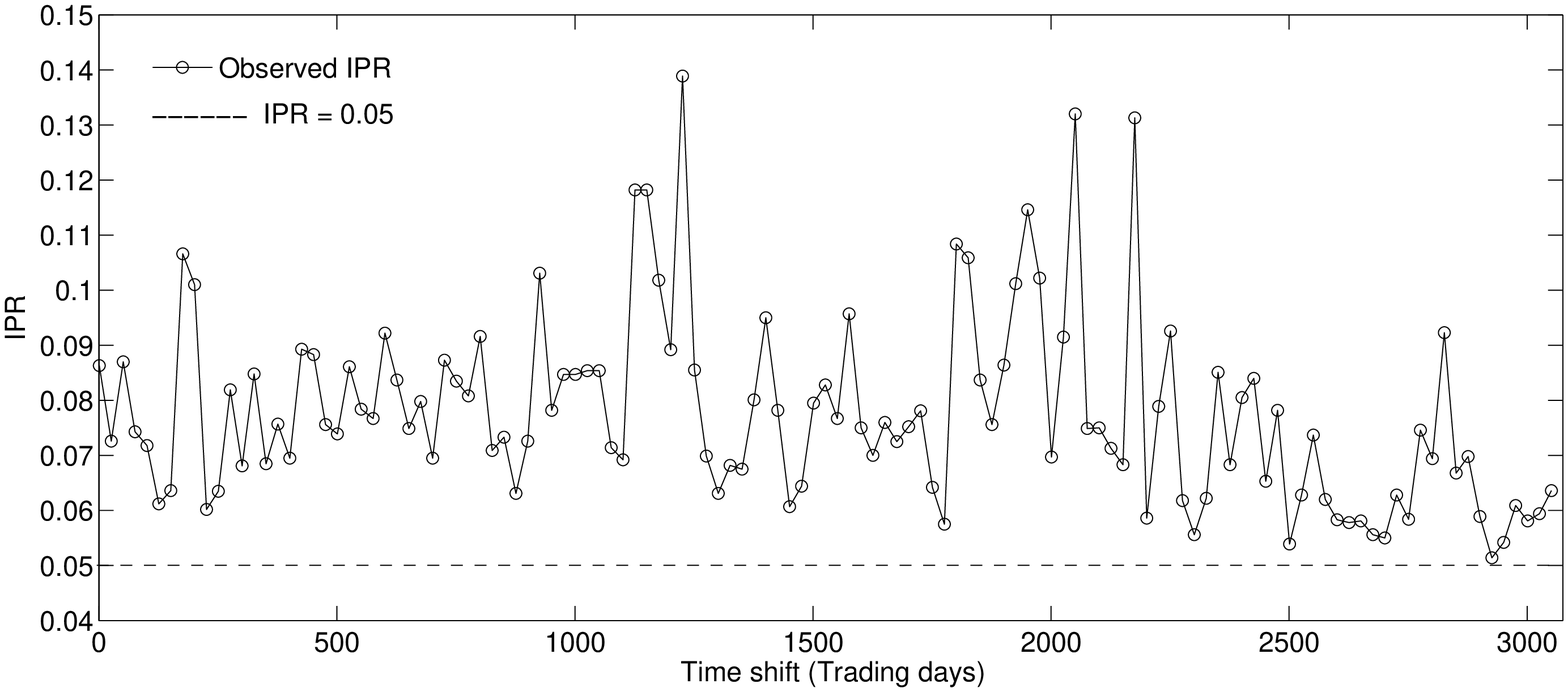}
\caption{IPR for the eigenvector $U^{20}$ as a function of time which is obtained from correlation matrix $C$ constructed from daily returns of 20 financial indices for 123 time windows of 25 days each. The dashed line marks the value 0.05 of IPR when all components contribute equally.}
\label{IPR20}
\end{figure}

\begin{figure}
\centering
\includegraphics[scale=0.4]{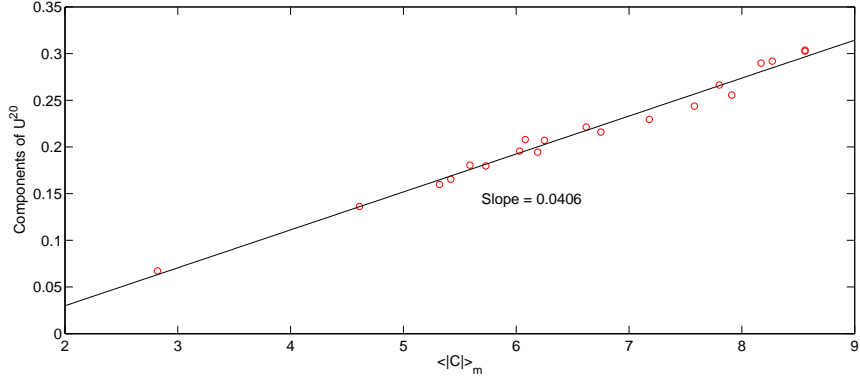}
\caption{Components of eigenvector $U^{20}$ corresponding to the largest eigenvalue with the extent
to which every individual index is correlated in the market, denoted by $<|C|>_{m}$. The line obtained by least square curve fitting has a slope =0.0406.}
\label{U20meanCforAllindices}
\end{figure}

\begin{figure}
\centering
\includegraphics[scale=0.4]{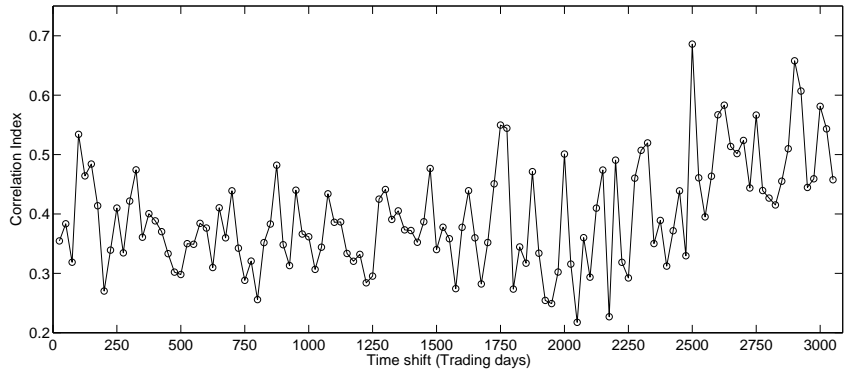}
\caption{Temporal evolution of the correlation index (CI) of the financial indices. The results are obtained from the
correlation matrix $C$ constructed from daily returns of 20 indices for 123 progressing time windows of 25 days each.}
\label{CIall25windows}
\end{figure}

\begin{figure}
\centering
\includegraphics[scale=0.4]{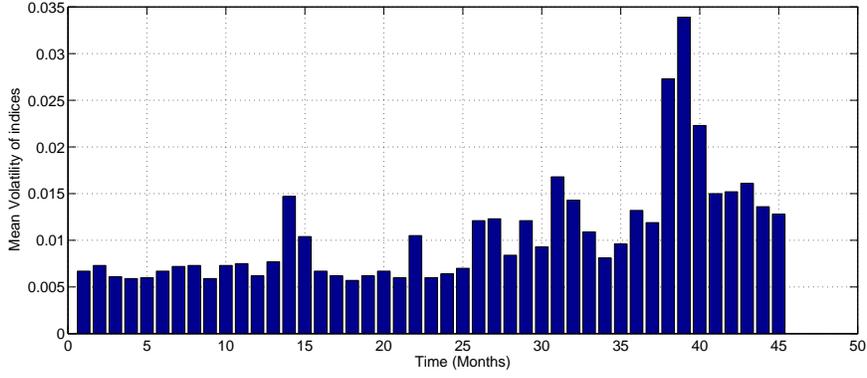}
\caption{Mean volatility of 20 financial indices. }
\label{meanvolatilitybar}
\end{figure}

\begin{figure}
\centering
\includegraphics[scale=0.4]{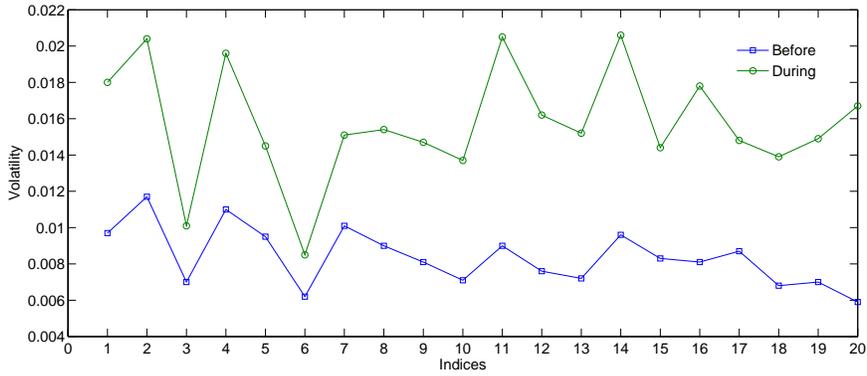}
\caption{volatility of 20 financial indices before and during the crisis.}
\label{volatility}
\end{figure}

\begin{figure}
\centering
\includegraphics[scale=0.4]{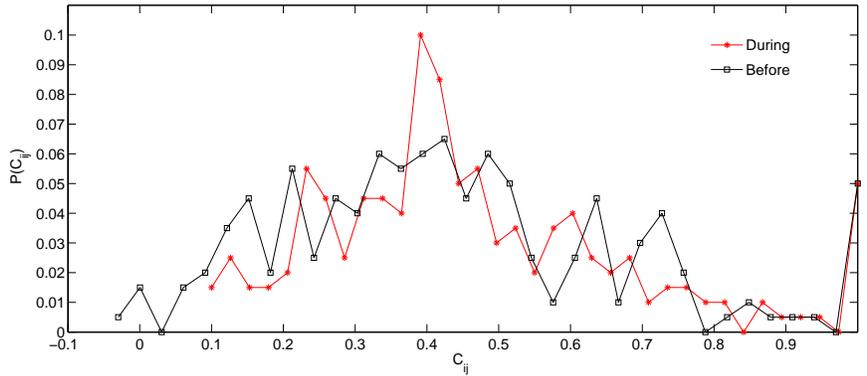}
\caption{Plot of the probability density of elements of correlation matrix C calculated using daily returns of 20 indices
before and during the crisis. We find the average magnitude of correlation $\langle|C| \rangle= 0.435$ before
and $\langle|C| \rangle= 0.463$ during the crisis respectively.}
\label{pcij}
\end{figure}

\begin{figure}
\centering
\includegraphics[scale=0.4]{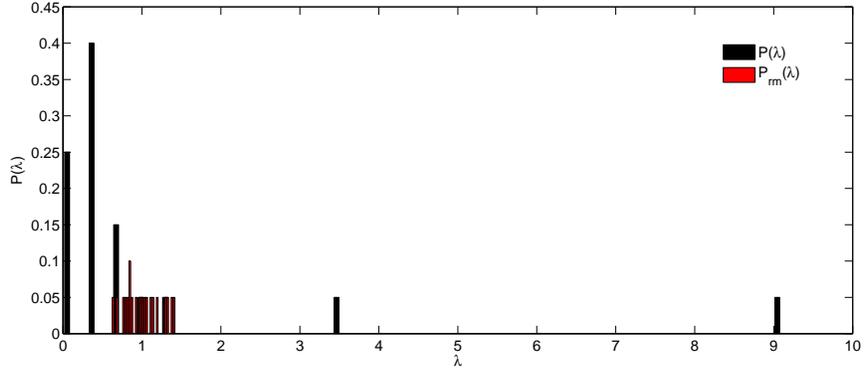}
\caption{Comparison of probability density function of 20 financial indices before the crisis. For N=20 indices, T=387 days and Q=19.35,
 $\lambda^{rand}_{min}=0.597$ and $\lambda^{rand}_{max}=1.506$ and $\lambda^{real}_{min}=0.0527$ and $\lambda^{real}_{max}=9.045$. }
\label{prmlamdabeforecrisis}
\end{figure}

\begin{figure}
\centering
\includegraphics[scale=0.4]{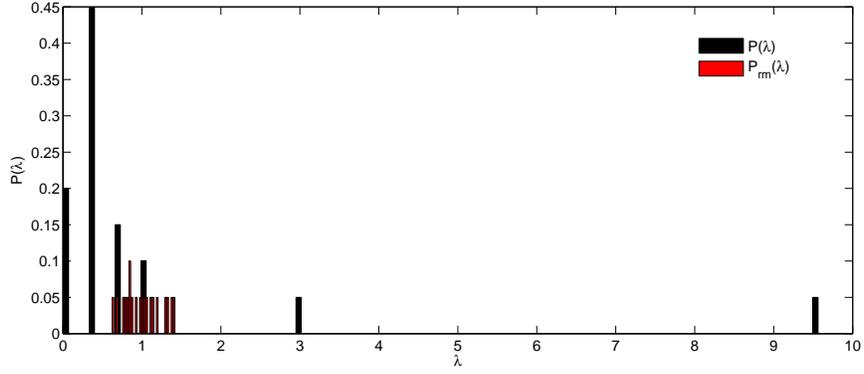}
\caption{Comparison of probability density function of 20 financial indices during the crisis. For N=20 indices, T=387 days and Q=19.35,
 $\lambda^{rand}_{min}=0.597$ and $\lambda^{rand}_{max}=1.506$ and $\lambda^{real}_{min}=0.0388$ and $\lambda^{real}_{max}=9.528$.}
\label{prmlamdaduringcrisis}
\end{figure}

\begin{figure}
\centering
\includegraphics[scale=0.4]{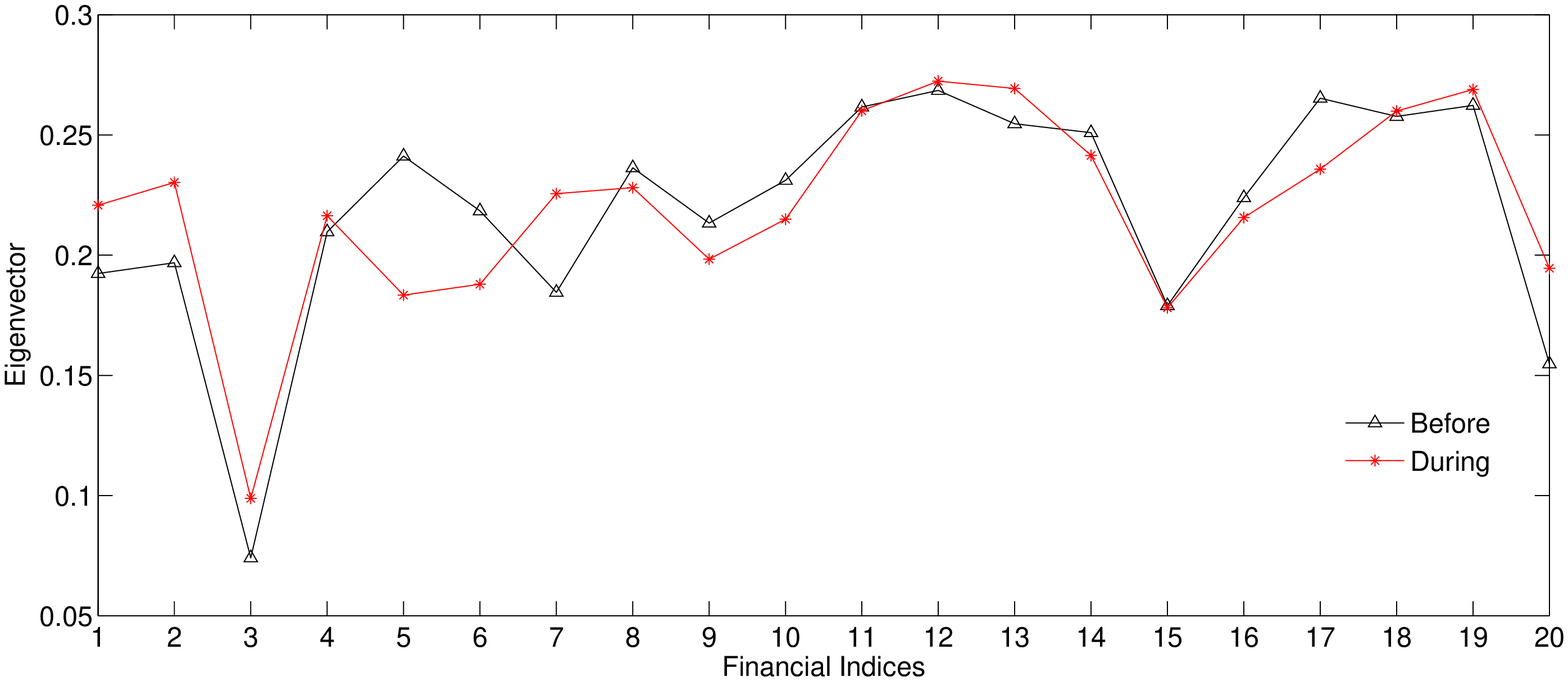}
\caption{Comparison of eigenvectors corresponding to first largest eigenvalue before and during the financial crisis of 2008
respectively. No significant difference is observed except the financial indices of Indonesia, Malaysia, and Mexico.}
\label{evectoroflargestev}
\end{figure}

\begin{figure}
\centering
\includegraphics[scale=0.4]{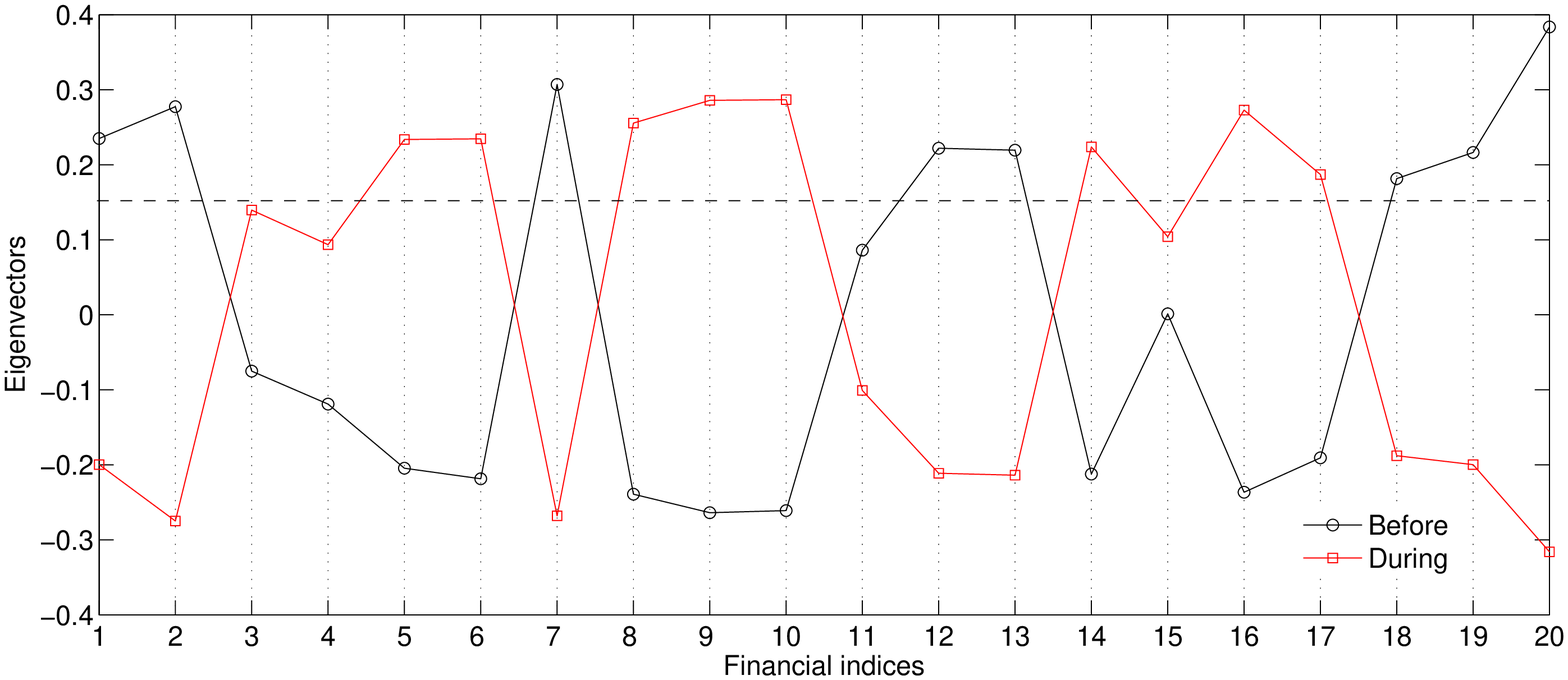}
\caption{Comparison of eigenvectors corresponding to second largest eigenvalue. Before crisis indices of Americas (Argentina, Brazil, Mexico, US) and Europe (France, Germany, Switzerland) contribute significantly while during the crisis Asia/Pacific (Indonesia, Malaysia, South Korea, Taiwan, Australia, Hong Kong, Japan, Singapore) contribute significantly. These sectors are
formed on the basis of geographical location.}
\label{eigenvectorscorresponsto2ndlargestevalue}
\end{figure}

\begin{figure}
\centering
\includegraphics[scale=0.4]{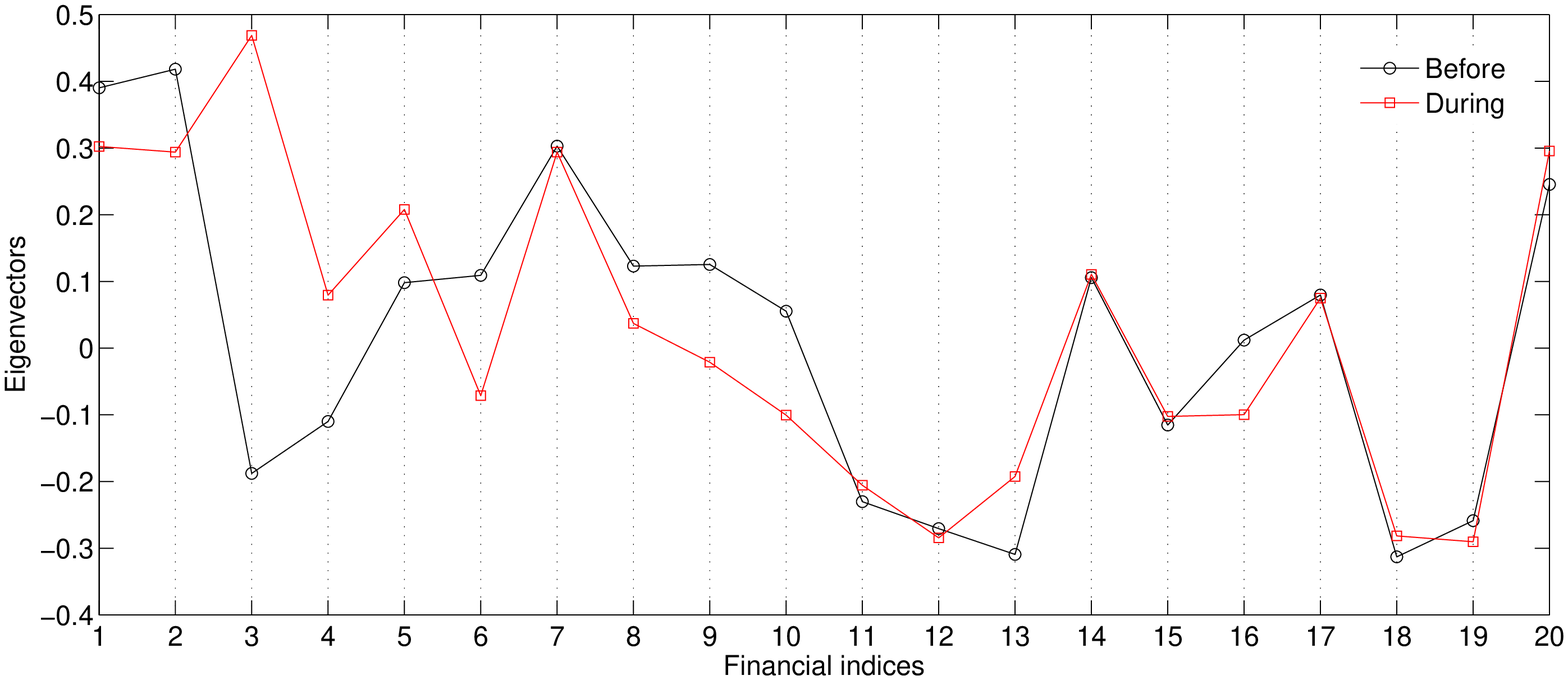}
\caption{Comparison of eigenvectors corresponding to third largest eigenvalue before and during the financial crisis of 2008
respectively.}
\label{eigenvectorscorresponsto3rdlargestevalue}
\end{figure}

Random matrix theory (RMT) originally developed \cite{mehta91} to study the interaction in complex quantum systems has been
useful in the analysis of universal and non-universal properties of cross-correlations between different
stocks \cite{gopistanely99,laloux99,plerou99,kulkarni07,sitabhra07}. Let $S{_i}(t)$ and $R{_i}(t)$ denote the daily closing prices and returns of indices $i$ at
time $t$ ($i=1,2,...,N ; t=1,2,...,L$), respectively. The logarithmic returns $R{_i}(t)$ can be defined as,\\
\begin{equation}
R{_i}(t) \equiv \ln (S{_i}(t+\Delta t)) - \ln(S{_i}(t)),
\label{eq.1}
\end{equation}
where $\Delta t=1$ day is the time lag.
The normalized returns for indices $i$ is defined as, \
\begin{equation}
r_i(t)\equiv {R_i(t) - \langle R_i \rangle \over \sigma_i}\,
\label{eq.2}
\end{equation}
where $\sigma_i \equiv \sqrt{\langle R_i^2 \rangle - \langle R_i
\rangle^2} $ is the standard deviation of $R_i$, and
$\langle\cdots\rangle$ denotes a time average over the period
studied. We then compute the equal-time cross-correlation matrix $C$ with elements,
\begin{equation}
C_{ij} \equiv \langle r_i(t) r_j(t) \rangle\,.
\label{eq.3}
\end{equation}
The elements of $C_{ij}$ are limited to the domain
$-1 \leq C_{ij} \leq 1$, where $C_{ij}=1$ defines perfect positive
correlations, $C_{ij}=-1$ corresponds to perfect negative correlations, and
$C_{ij}=0$ corresponds to no correlation.\
If $N$ time series of length $T$ are mutually uncorrelated, the resulting cross-correlation
matrix is termed as a Wishart matrix. Statistical properties of such random matrices
are known ~\cite{bowick91}. In the limit of $N \rightarrow \infty \,, L \rightarrow \infty$,
such that $Q\equiv L/N\geq1$, the probability distribution $P_{rm}(\lambda)$ of the eigenvalue $\lambda$
is given by,
\begin{equation}
P_{rm}(\lambda) = {Q \over 2 \pi}\, { \sqrt{(\lambda^{rand}_{max} - \lambda)(\lambda - \lambda^{rand}_{min})} \over \lambda}\, \,\,,
\label{densuncorr}
\end{equation}
for $\lambda$ within the bounds $\lambda^{rand}_{min}\leq \lambda_i \leq \lambda^{rand}_{max}$, where $\lambda^{rand}_{min}$ ($\lambda^{rand}_{max}$) are the lower (upper) bound given by,
\begin{equation}
\lambda^{ran}_{max(min)} = [1\pm(1/\sqrt Q)]^{2}.
\label{lambdamaxmin}
\end{equation}
We study the data of $N=20$ financial indices before and during the financial crisis of 2008 \cite{stanley10,bpstanley10}.
The volatility gives us a measure of the market fluctuations. We quantify the volatility, as the local average of the absolute
value of daily returns of indices in an appropriate time window of T days, as an estimate of volatility in that period
$v(t) = {\displaystyle\sum_{t=1}^{T-1} |R(t)| \over T-1}$. We compute the mean volatility of all indices (June 7, 2007 to
November 30, 2009) by taking T=25 days which is shown in Fig.~\ref{meanvolatilitybar}. The volatility for two periods
June 7, 2006 to November 30, 2007 and December, 2007 to June, 2009 for individual countries is shown in Fig.~\ref{volatility},
we consider these two periods as the period before and during the financial crisis of 2008 respectively.
We then construct the cross-correlation matrix $C_{ij}$ from daily returns of N=20 indices before and during crisis periods.
The probability densities of $C_{ij}$, $P(C_{ij})$ for both periods are compared in Fig.~\ref{pcij}.
The largest eigenvalue deviating from RMT prediction reflects that some influence of the full global market is common to all
indices and it alone yields "genuine" information hidden in $C$. The range of eigenvalues within the RMT bounds corresponds to
noise and do not yield any system specific information. Therefore, we compare the properties of $C$ with those of a random
correlation matrix in Fig.~\ref{prmlamdabeforecrisis} and Fig.~\ref{prmlamdaduringcrisis} respectively to extract information about the cross correlations.
If there is no correlation between these financial indices, the eigenvalues should be
bounded between RMT predictions i.e. $\lambda^{rand}_{min}=0.597$ and $\lambda^{rand}_{max}=1.5063$. We find that
before the financial crisis period (June 7, 2006 to November 30, 2007), $\lambda^{real}_{min}=0.0527$ and
$\lambda^{real}_{max}=9.0454$; during financial crisis period (December, 2007 to June, 2009), $\lambda^{real}_{min}=0.0388$
and $\lambda^{real}_{max}=9.5282$. Here, we find that largest eigenvalues deviate significantly from the upper bound
$\lambda^{rand}_{max}$ which shows a strong correlation between financial indices. We also find an increase
in the value of $<C_{ij}>=0.4353$ before the crisis and $<C_{ij}>=0.4634$ during the crisis period. Since the largest
eigenvalue represents the collective information about the correlation between different indices therefore we expect its
trend to be dependent on the market conditions \cite{laloux99,plerou99,gobaix03,shen09} and can be seen in
Fig.~\ref{evectoroflargestev} which is plotted for eigenvectors corresponding to first largest eigenvalue. We find that eigenvectors corresponding to second largest eigenvalue give the information about a sector formation in global
financial indices. In Fig.~\ref{eigenvectorscorresponsto2ndlargestevalue}, we compare eigenvectors corresponding to second
largest eigenvalue before and during the financial crisis. Countries corresponding to financial indices above eigenvector
threshold $0.15$ that are contributing most to eigenvectors corresponding to second largest eigenvalues are as follows: Argentina, Brazil, Mexico, France, Germany, Switzerland, UK, US (before the crisis) and Indonesia, Malaysia, South Korea, Taiwan, Australia, Hong Kong, Japan, Singapore (during the crisis). We find that these sectors are forming on the basis of the geographical location. Before crisis indices of Americas (Argentina, Brazil, Mexico, US) and Europe (France, Germany,
Switzerland) contribute significantly while during the crisis indices of Asia/Pecific (Indonesia, Malaysia, South Korea, Taiwan, Australia, Hong Kong, Japan, Singapore) contribute significantly to the eigenvectors corresponding to second largest eigenvalue. The classification of major world indices has been considered as \cite{yahoo}. However, eigenvectors corresponding to third largest eigenvalue (Fig.~\ref{eigenvectorscorresponsto3rdlargestevalue}) does not give so much information as it
is near the random matrix bound. We also analyze the eigenvalue dynamics of correlation matrices $C$ constructed by using $3088$ daily returns of $20$ indices using a sliding window of 25 days. The daily closing prices and logarithmic returns of 20 financial indices are shown in Fig.~\ref{pricesret}. The correlation matrix was constructed from 20 financial indices having the 3088 returns. Fig.~\ref{largestev} shows the trend of first, second, and third largest eigenvalue over each of these sliding windows. Here, we find increase in the first and second largest eigenvalues during the financial crisis of 2008 while third largest eigenvalues do not show significant variation. The dynamics of smallest eigenvalues are shown in Fig.~\ref{smallestev}. We do not observe any significant pattern.\
We also analyze the evolution of the structure of the last eigenstate, $U^{20}$ by evaluating the Inverse Participation
Ratio (IPR) which allows quantification of the number of components that participate significantly in each eigenvector and
tells us more about the level and nature of deviation from RMT. The IPR of the eigenvector $u^k$ is defined
by $ I^k \equiv \sum_{l=1}^N\, [u^k_{l}]~^4\,$,
where $u^k_{l}$, $l=1,\dots,N$ are the components of eigenvector $u^k$. Thus IPR allows us to compute the inverse of the
number of eigenvector components that contribute significantly to each eigenvector. Fig.~\ref{IPR20} shows the IPR of 20
financial indices and is closest to 0.05 (=1/20), the value we would expect when all components contribute equally, in the
most volatile periods of time span. This has similar characteristics to those found for different
indices.\
The average magnitude of correlations of returns of every index m with all indices
 $n=1,2,...,N$ is $<|C|>_{m} = \frac{1}{N-1}\sum_{k=1}^N|C_{mk}|,$ when $m \neq k$. The
variation of $<|C|>_{m}$ for $m = 1,2,...,N$ with the corresponding components of $U^{20}$
in Fig. \ref{U20meanCforAllindices} shows a strong linear positive relationship between
the two at all times. Now define a projection vector $S$ with elements $S_{m} = <|C|>_{m}$, where $m=1,2,...,20,$. We
obtained a quantity $X_{m}(t)$ by multiplying each element $S_{m}$ by the square of the corresponding component of $U^{20}$
for each time window $t$, $X_{m}(t)=(U_m^{20})^2S_m$, where $m=1,2,...,20$. The idea behind this is to weigh the average
correlation possessed by every index $m$ in the market according to the contribution of the corresponding component to
the last eigenvector $U^{20}$ \cite{gopihes01}, thereby neglecting the contribution of less significant participants
(the one negligible in magnitude) in $U^{20}$. The quantity $X$ in some sense represents the effective magnitude of
correlation of financial indices. The sum of correlation of magnitudes is obtained as $CI(t)=\sum_{m=1}^{20}X_{m}(t),$ at
time $t$ and may be expected to reflect the correlation of the market at that time. This is called the Correlation
Index (CI) \cite{gopihes01,kulkarni07}. Temporal evolution of the correlation index is shown in Fig. \ref{CIall25windows}.
\section{Construction and analysis of the correlation network of financial indices}
\begin{figure}
\fbox{(a)\includegraphics[scale=0.2]{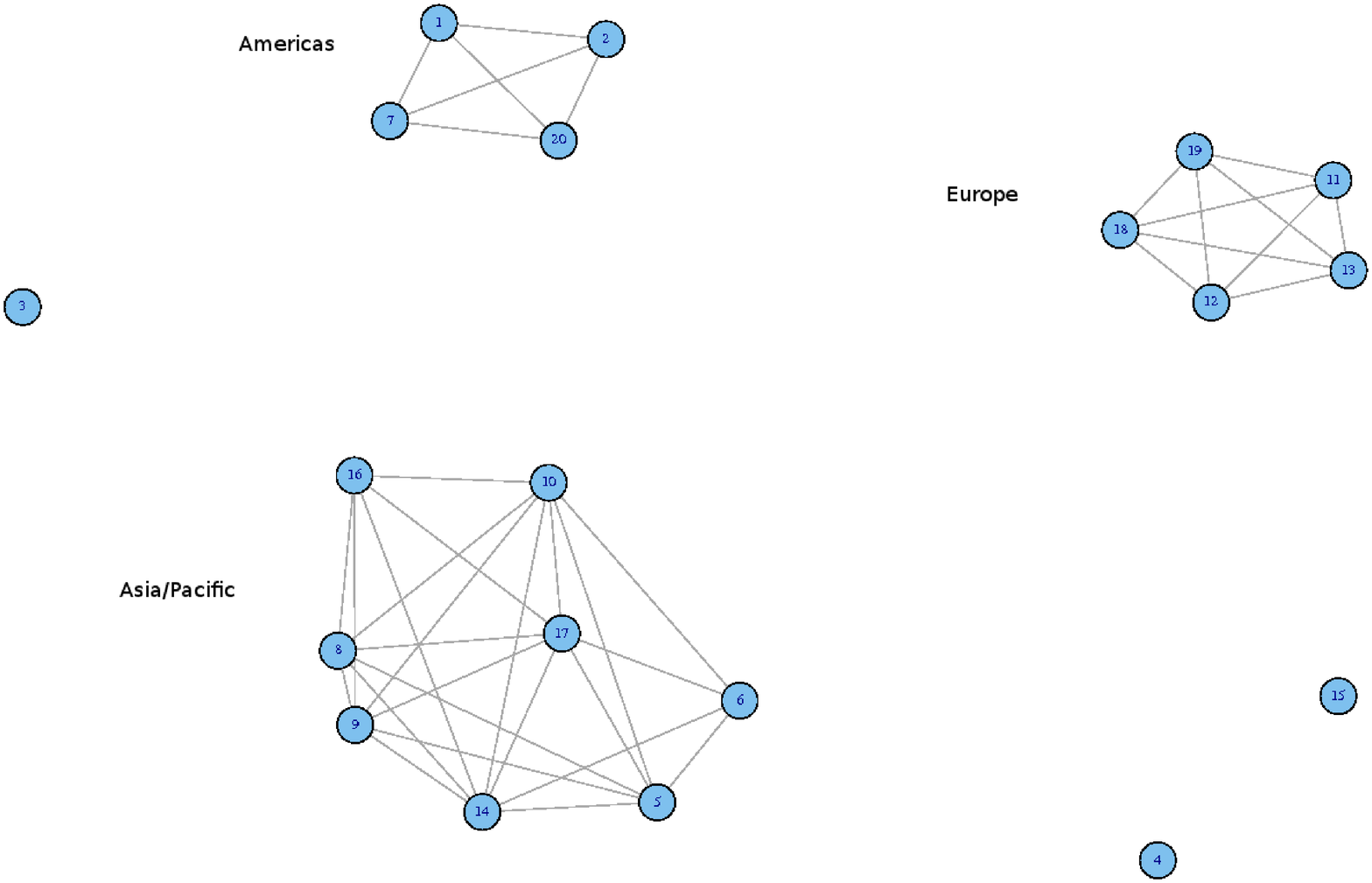}}
\fbox{(b)\includegraphics[scale=0.2]{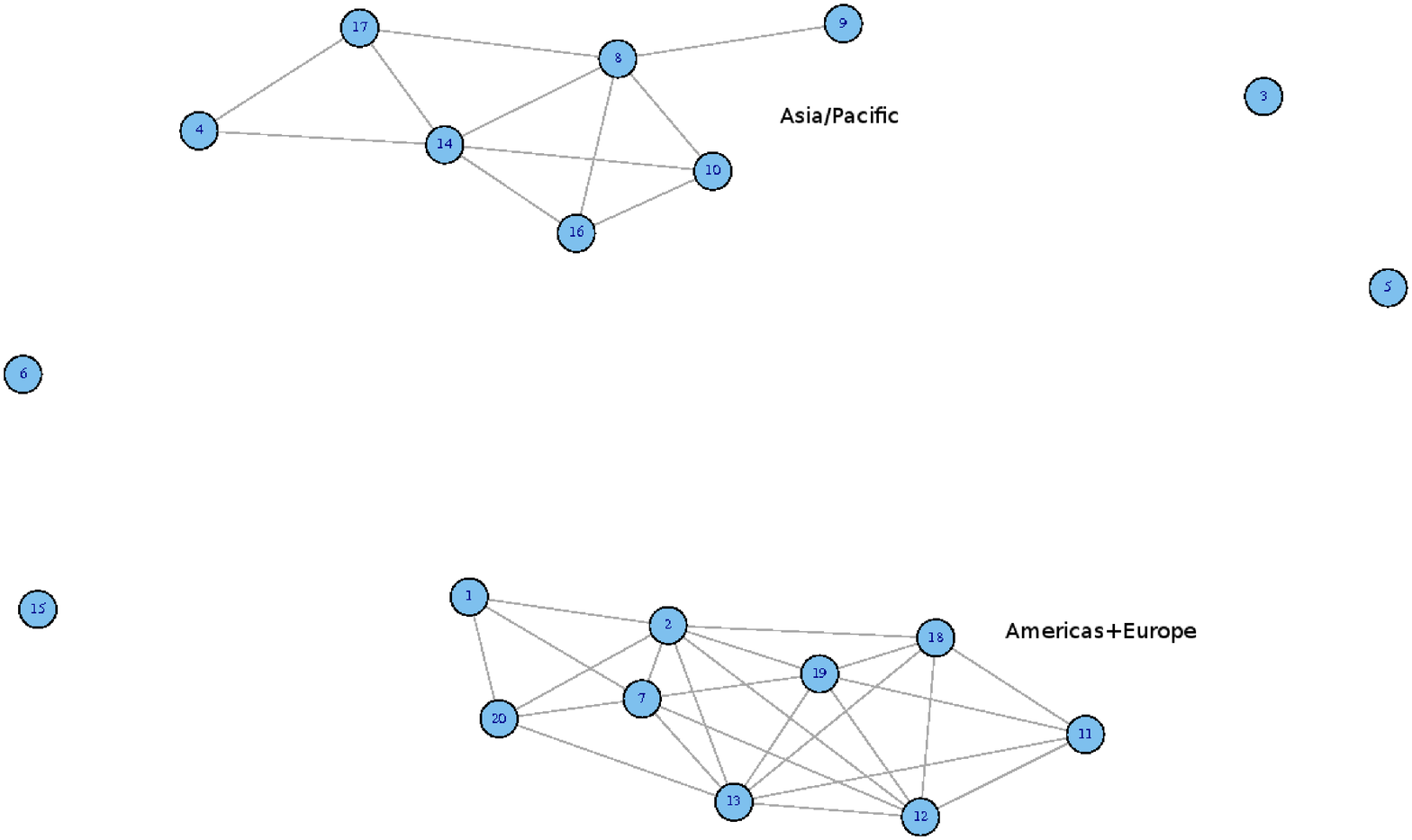}}
\fbox{(c)\includegraphics[scale=0.2]{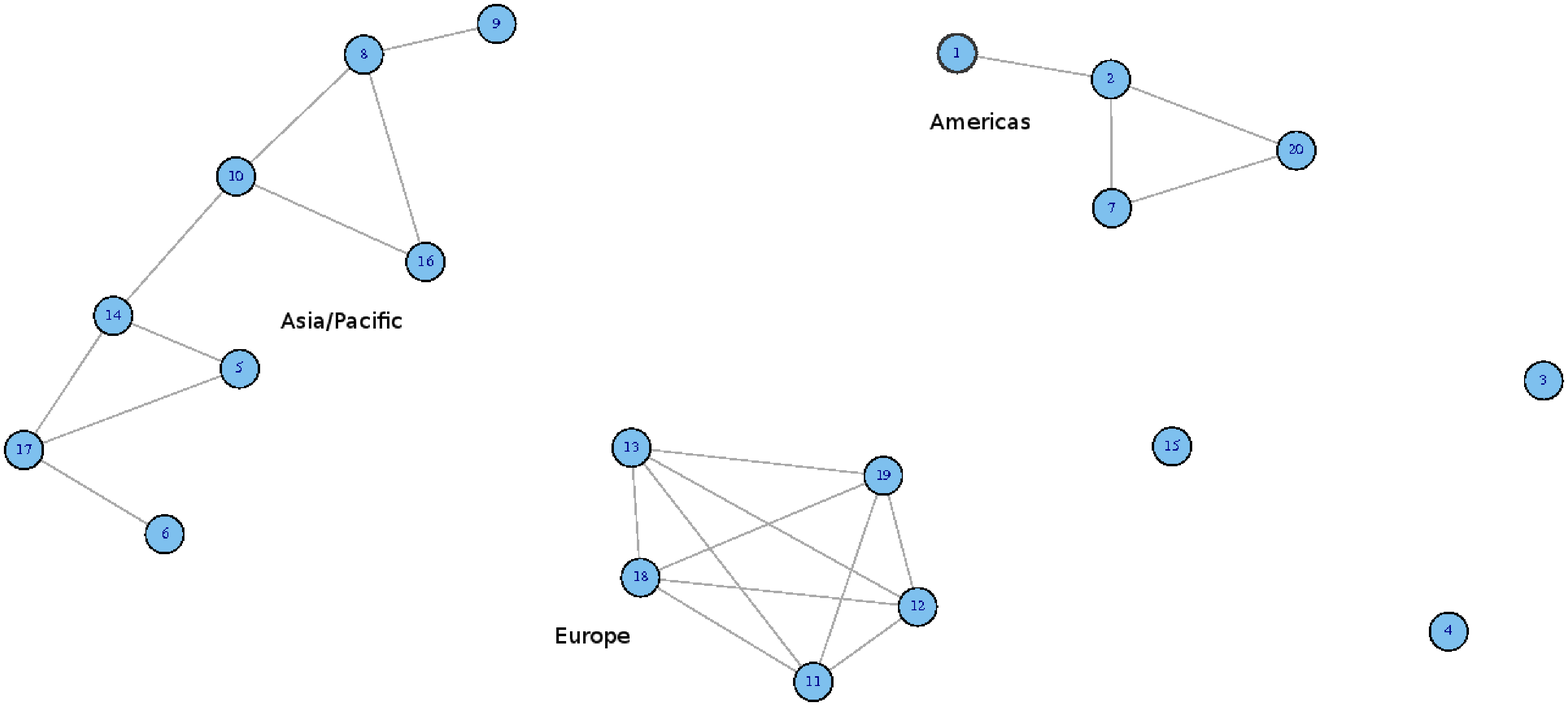}}
\fbox{(d)\includegraphics[scale=0.2]{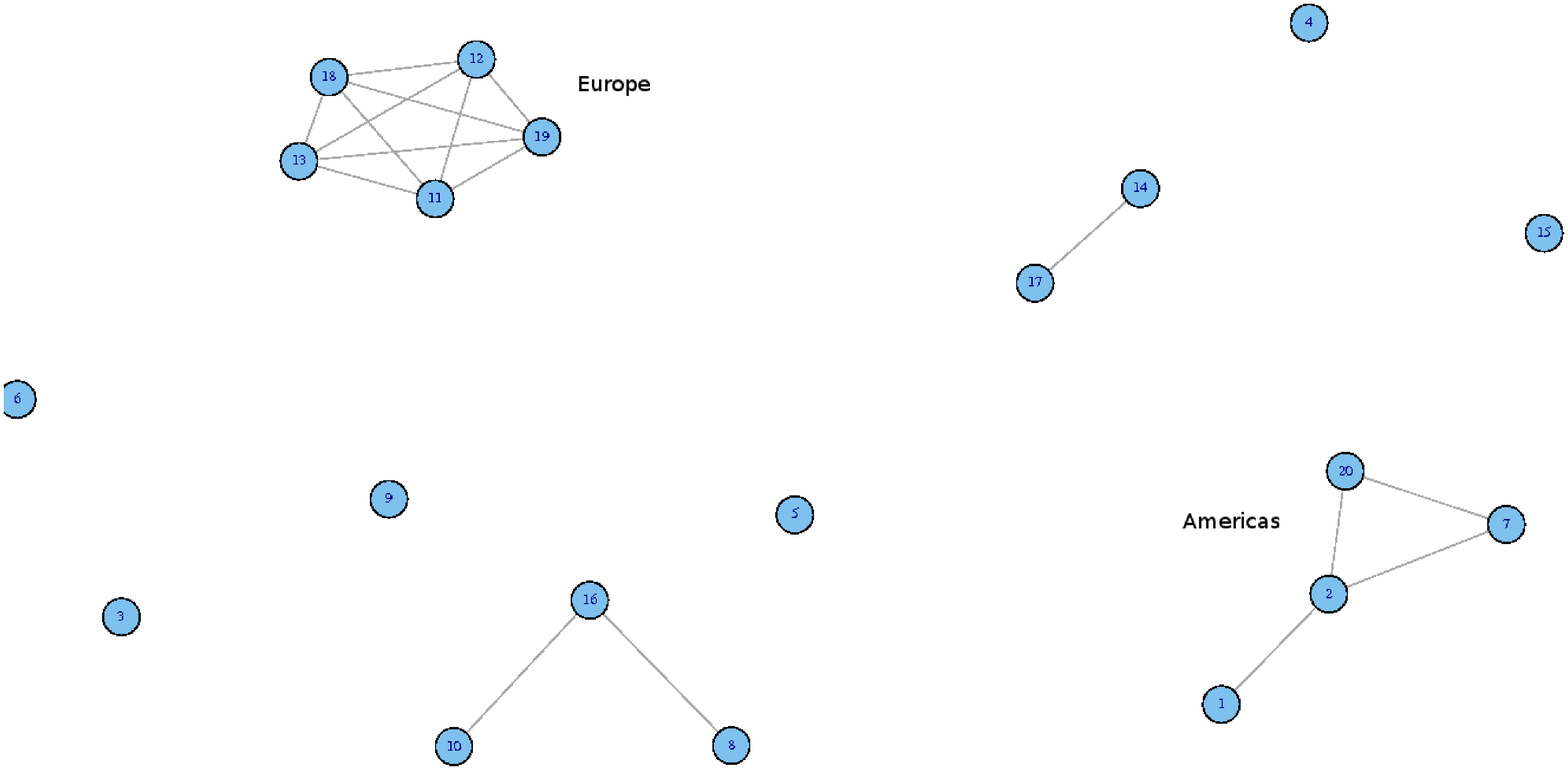}}
\fbox{(e)\includegraphics[scale=0.2]{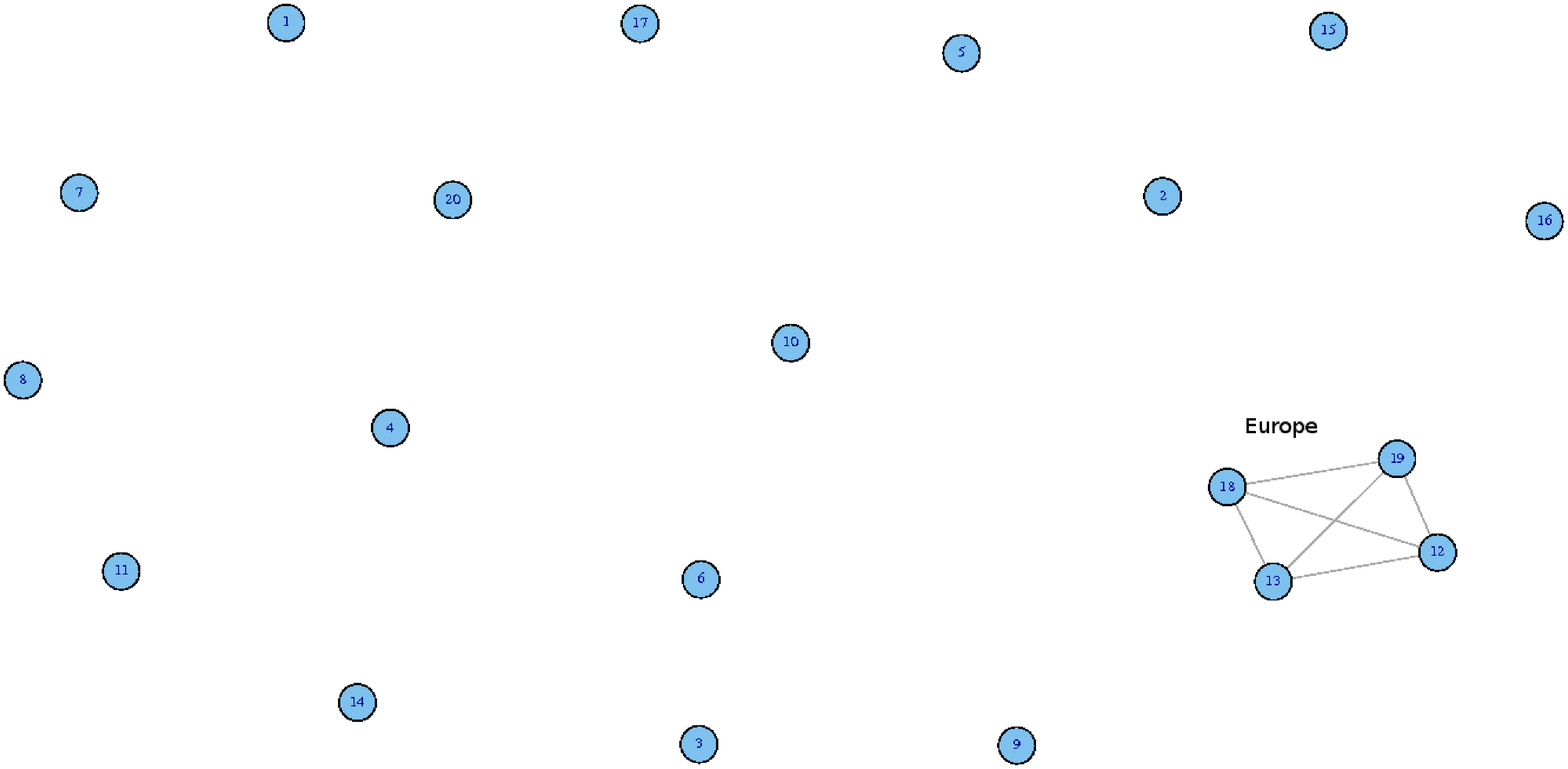}}
\fbox{(f)\includegraphics[scale=0.2]{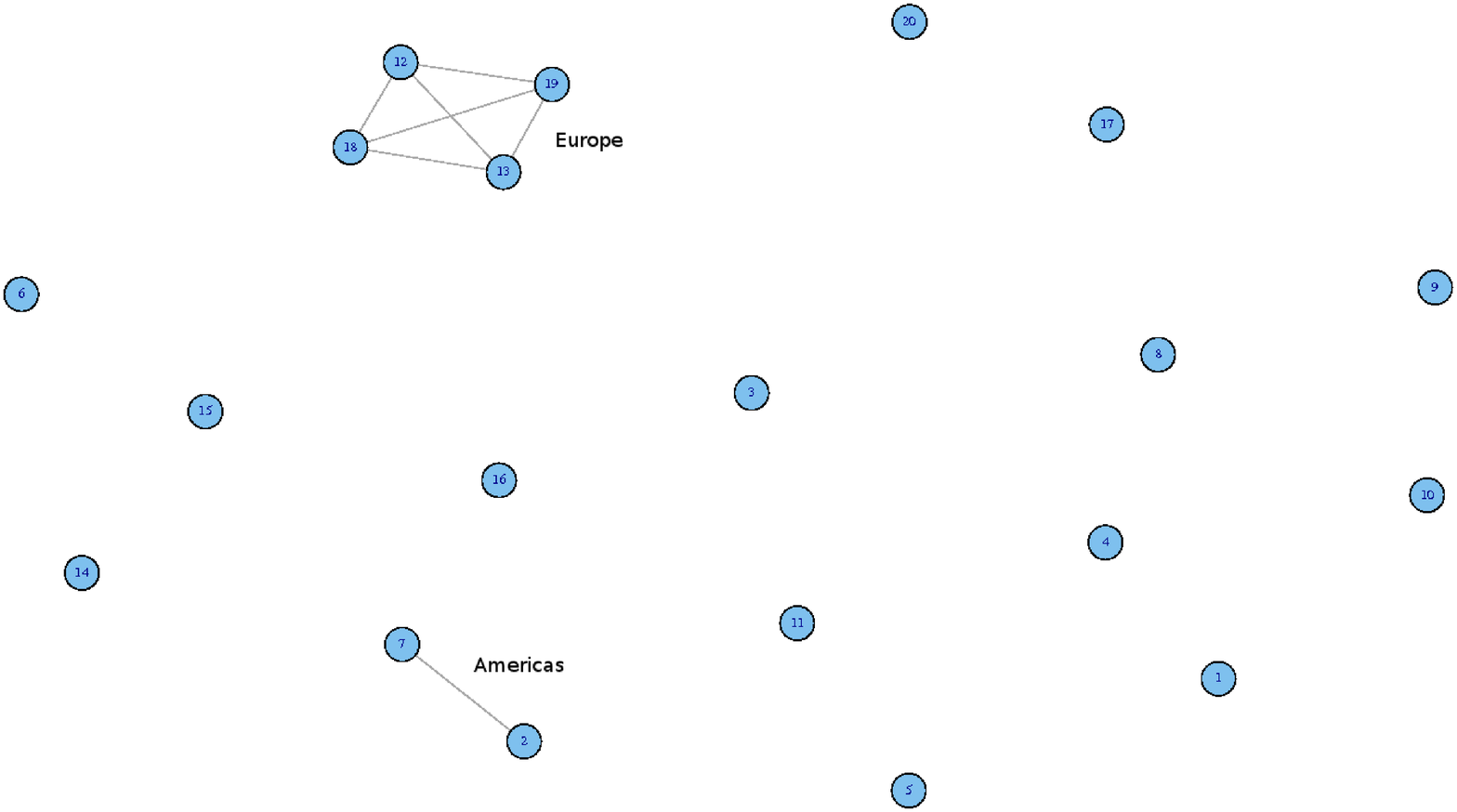}}
\fbox{(g)\includegraphics[scale=0.2]{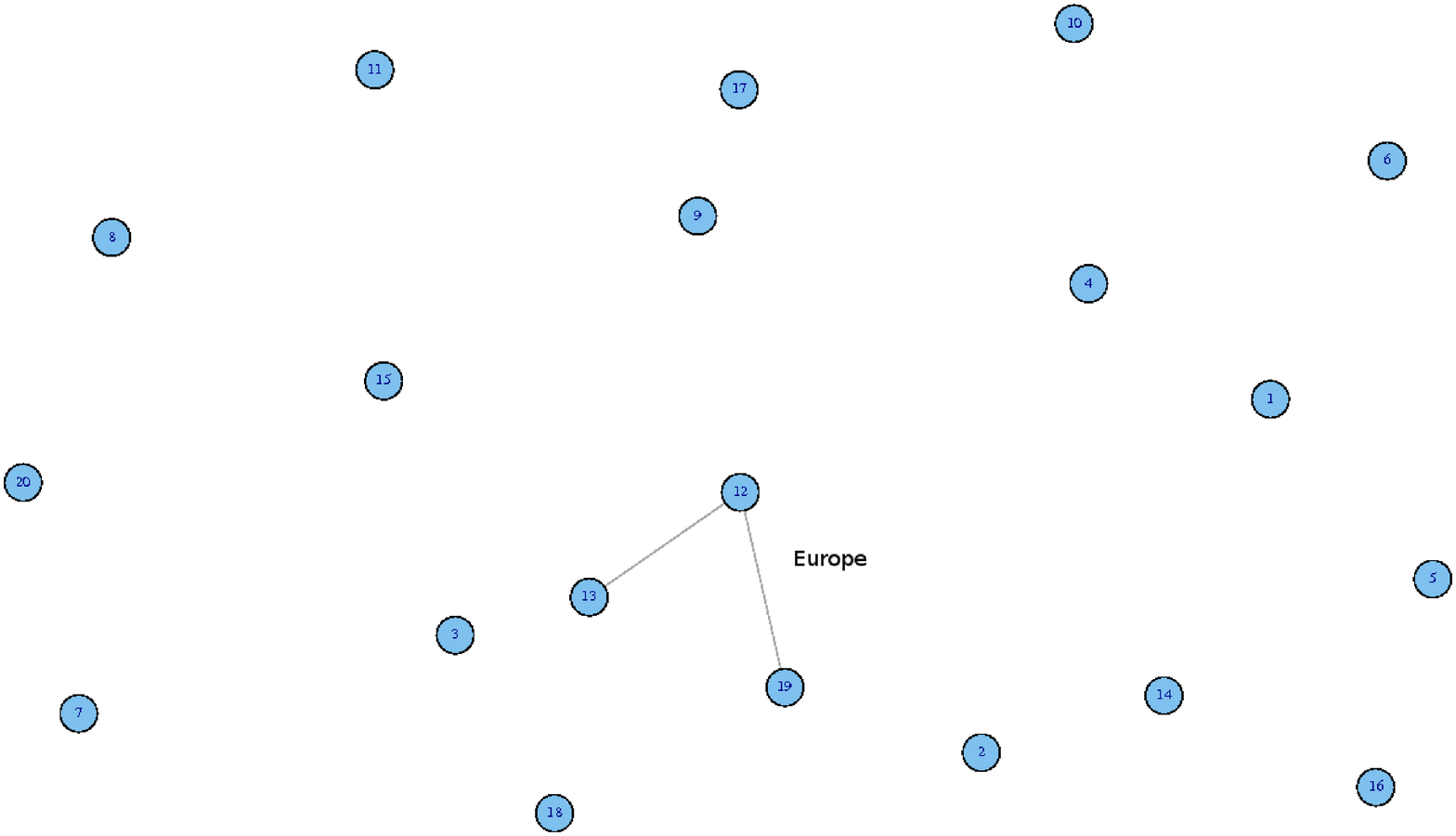}}
\fbox{(h)\includegraphics[scale=0.2]{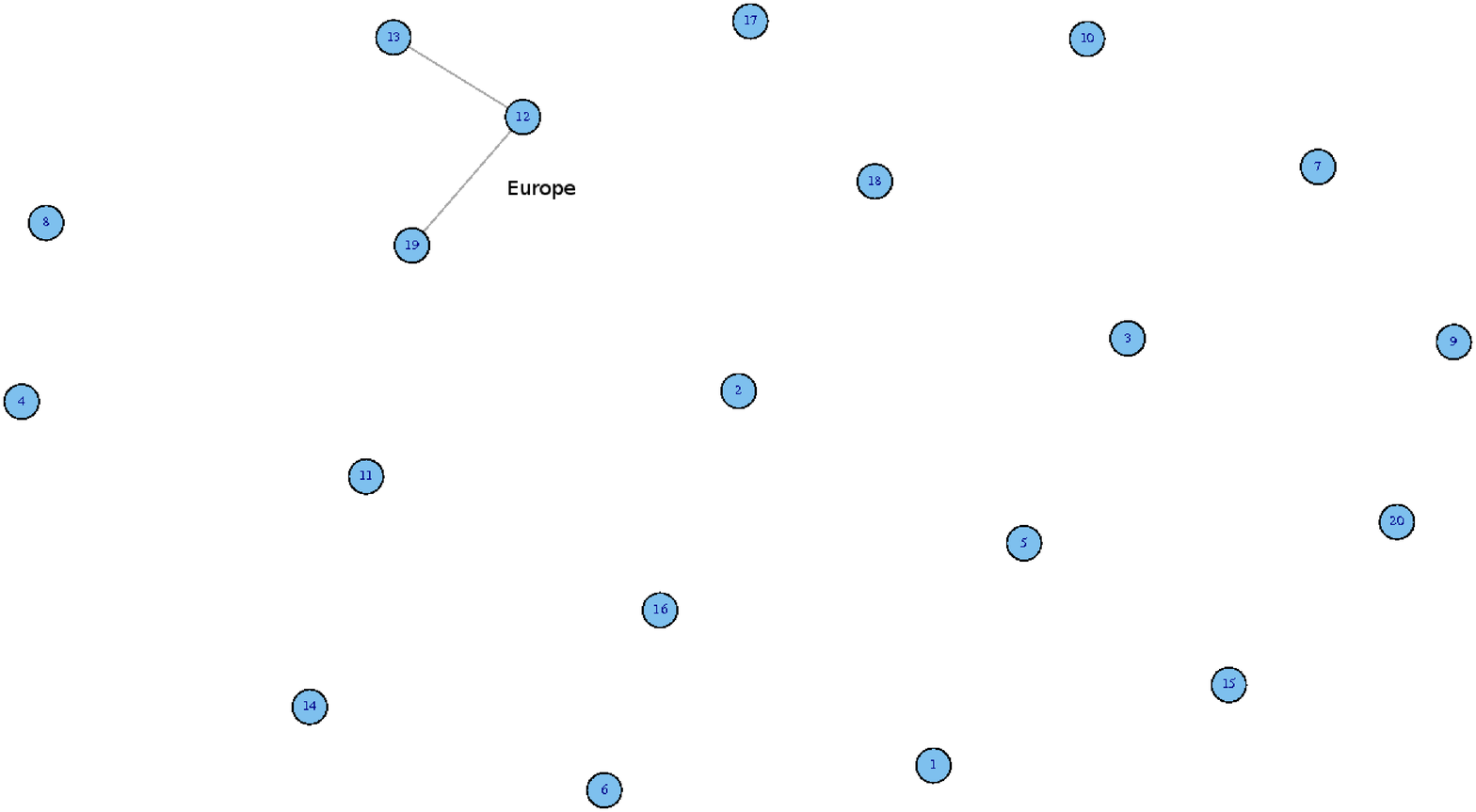}}
\caption{The financial network of 20 indices at different threshold $\theta$ before and during the crisis.
(a) $\theta=0.6$ (before): Cluster of financial indices of, \textbf{Americas} (Argentina, Brazil, Mexico, and
US), \textbf{Europe} (Austria, France, Germany, Switzerland, and UK), and \textbf{Asia/Pecific} (Australia, Hong Kong, Indonesia, Malaysia, Japan, Singapore, South Korea, Taiwan). (b) $\theta=0.6$ (during): Clusters of indices of \textbf{Asia/Pecific} (Australia, Hong Kong, Japan, India, Singapore, South Korea, Taiwan), \textbf{Americas} (Argentina, Brazil,
Mexico,US), and \textbf{Europe} (Austria, France, Germany, Switzerland, UK). (c) $\theta=0.7$ (before) (d) $\theta=0.7$ (during)
(e) $\theta=0.8$ (before) (f) $\theta=0.8$ (during). (g) $\theta=0.9$ (before) (h) $\theta=0.9$ (during). At $\theta=0.9$ indices corresponding to Europe: France, Germany and UK consistently constitute the most tightly linked markets both before and during the crisis.}
\label{thetap6to9}
\end{figure}

The main idea of constructing the index correlation network is as follows: Let the set of index represent the set of vertices
of the network. A certain threshold $\theta$ is specified such that $-1 \leq \theta\leq 1$ and an undirected edge connecting the
vertices i and j if the correlation coefficient $C_{ij}$ is greater than or equal to $\theta$. Different values of $\theta$
define the networks with the same set of vertices, but different set of edges \cite{huang2009}.
Let the graph $G = (V,E)$ represent the index correlation network, where V and E are the set of vertices edges respectively.
E is defined by
\[
E =
 \left\{
  \begin{array}{ll}
 e_{ij}=1,  & \mbox{$i\neq j$ and $C_{ij}\geq\theta$} \\
 e_{ij}=0,  & \mbox{$i=j$}.
  \end{array}
 \right\}
\]
We construct networks for different values of threshold $\theta$ in the range 0 to 0.9. We find that at threshold $\theta=0.2$ the network is fully connected. In the network at threshold $\theta=0.6$ (Fig. ~\ref{thetap6to9}) the Americas, Europe and Asia/pacific forms different clusters before the crisis but during the crisis Americas and Europe forms a combined cluster of strong link between them. If we further increase the threshold $\theta$ up to 0.9 we find that European countries: France, Germany and UK, consistently constitute the most tightly linked markets for both before and during the crisis.
\subsection{Topology structure of the network}
\begin{figure}
\centering
\includegraphics[scale=0.4]{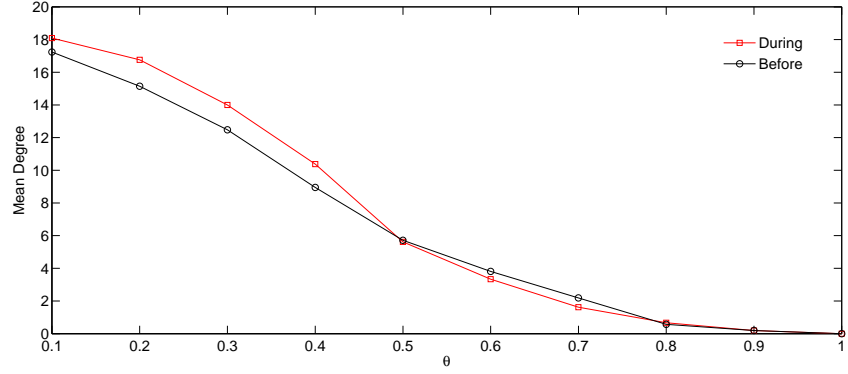}
\caption{Mean degree for various thresholds before and during the crisis.}
\label{MeanDegreetheta}
\end{figure}

\begin{figure}
\centering
\includegraphics[scale=0.4]{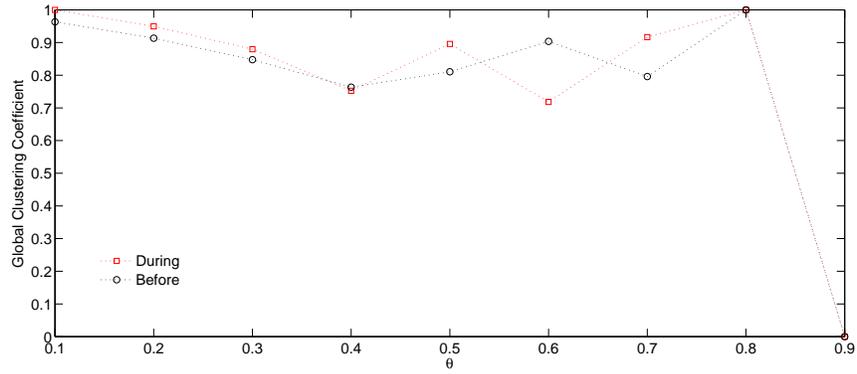}
\caption{Global clustering coefficients for various thresholds before and during the crisis. At $\theta=0.9$ there is no
triangle formation in the correlation network and there is only one triplet so its clustering coefficient is zero.}
\label{GlobalClusteringCoefftheta}
\end{figure}

\begin{figure}
\centering
\includegraphics[scale=0.4]{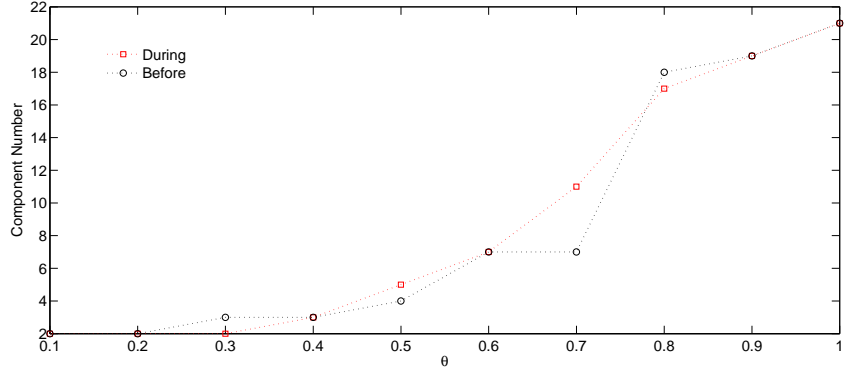}
\caption{Component number in the index correlation network under different correlation thresholds. When $\theta>0.9$, vertices
are nearly all isolated and the component number is approximately the vertex number. When $\theta\leq0.2$, networks are fully
connected and the component number is just 2.}
\label{componentNumbertheta}
\end{figure}

\begin{figure}
\centering
\includegraphics[scale=0.4]{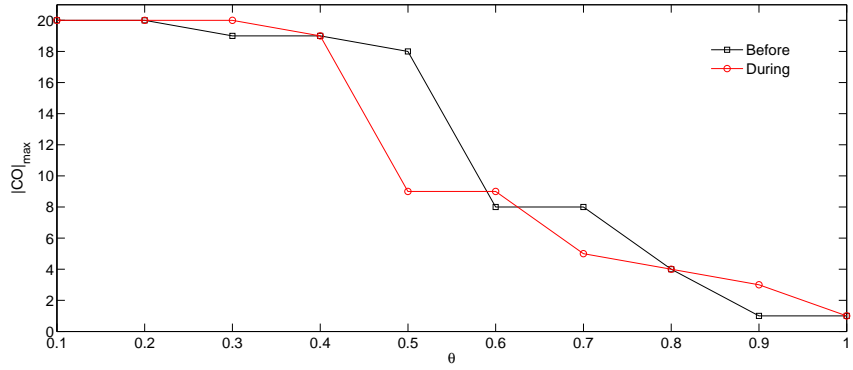}
\caption{Maximum connected component size $|CO|_{max}$ of index correlation network under different threshold. $\theta=0.2$
is a critical point when the financial correlation network is fully connected.}
\label{MaxCompSizetheta}
\end{figure}

\begin{figure}
\centering
\includegraphics[scale=0.4]{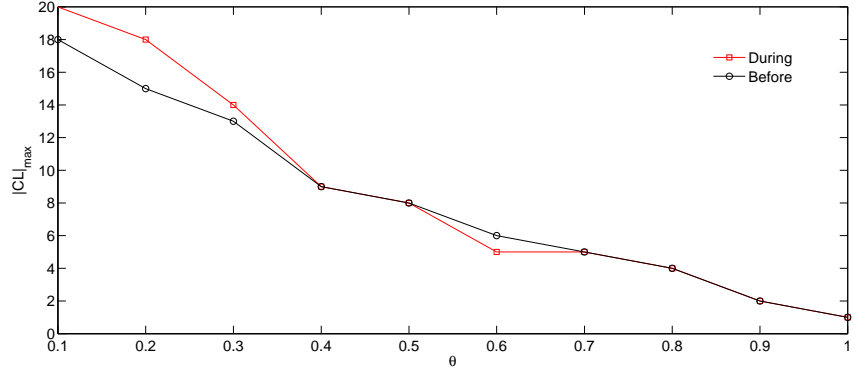}
\caption{If $\theta\geq0$, all indices in a clique are positively correlated with each other. In such a case a price fluctuation
of any one index will make all other indices price in this clique fluctuate towards the same direction.}
\label{maxCliqueSizetheta}
\end{figure}

\textbf{Mean Degree}: The degree of vertex $i$ is $k_{i}=\sum_{j\neq i}e_{ij}$ which denotes the vertex number
connecting with $i$. The mean degree is based upon the degree. It shows how many neighbors a node in the network has in average. This measure can only be calculated when the network has at least one edge connecting the nodes. In Fig.~\ref{MeanDegreetheta} we find that mean degree decreases as threshold increase because the number of connected vertices decreases with increase in the threshold. In such a network, most of the vertices have small degree and a small number of vertices have large degree. We call the later as "Hub" vertices. Especially, in the index correlation network, index $i$ having a large degree means that
it is correlated with many other indices in the sense of price fluctuation. So we can dig out the very indices that can most
accurately reflect market behavior by the degree of indices.\\
\textbf{Clustering coefficient}: If $k_{i}$ nearest neighbors of vertex $i$ have $m_{i}$ edges among them, the ratio of $m_{i}$ to $k_{i}(k_{i}-1)/2$ is the clustering coefficient of vertex $i$. The network clustering coefficient is calculated by averaging through the clustering coefficient of all vertices. The phenomenon of large clustering in actual networks motivates the appearance of small-world network models \cite{watts98, newman99, barrat00}. Specially, the network clustering coefficient of a index correlation network indicates the clustering property of indices in the meaning of price fluctuation correlation. Global clustering coefficients that is simply the ratio of the triangles and connected triples in the network are shown in Fig.~\ref{GlobalClusteringCoefftheta} for different correlation thresholds. It can be seen that values of clustering coefficients of index correlation network become smaller with increase in threshold up to $\theta=0.4$. At $\theta=0.9$ there is no triangle formation in the correlation network and there is only one triplet so its clustering coefficient is zero.\\
\textbf{Connected components}: The graph $G=(V,E)$ is connected if there is a path from any vertex to any vertex in the
set $V$. If the graph is disconnected, it can be decomposed into several connected subgraphs, which are referred to as
connected components \cite{erdos1960} of G. The size of a connected component $|CO|$ is defined as the vertex number in it.
The maximum component size is shown in Fig.~\ref{MaxCompSizetheta} for different thresholds.
In the index correlation network the component number represent financial indices that are correlated with each other. The
component number for various threshold are shown in Fig.~\ref{componentNumbertheta} before and during the crisis. We find component number depends on the value of correlation thresholds, it increases with increase in correlation threshold. The structural characteristics of components in the index correlation network indicate the correlation modes of indices in the global financial market. It can be seen in Fig.~\ref{componentNumbertheta} and Fig.~\ref{MaxCompSizetheta} that larger the component number, smaller is the maximum component size. For $\theta\leq0.2$ the network is fully connected. If we further increase
the threshold the network starts forming subclusters based on geographical location.\\
\textbf{Clique}: Given a subset $S\subseteq V$, by $G(S)$ we denote the subgraph induced by S. A subset $C\subseteq V$ is a
clique if $G(C)$ is a complete graph, i.e. it has all possible edges. The size of a clique $|CL|$ could be denoted by the
vertex number in it. The maximum clique problem is to find the largest clique in a graph. The financial interpretation of
the clique in the index correlation network is that it defines the set of indices whose price fluctuations exhibit a
similar behavior. In Fig~\ref{maxCliqueSizetheta} we find that maximum clique in the index correlation network decreases with increase in the value of threshold . We find that up to $\theta=0.4$ the maximum clique size is much larger during the crisis period and behaves the same in both period when $\theta\geq0.4$ except $\theta=0.6$, at which indices forms three different major sectors of Americas, Europe and Asia/Pecific before the crisis, but during the crisis financial indices of Americas and Europe combined and form a strong cluster.
\subsection{Minimum Spanning Tree}
\begin{figure}
\fbox{(a)\includegraphics[scale=0.4]{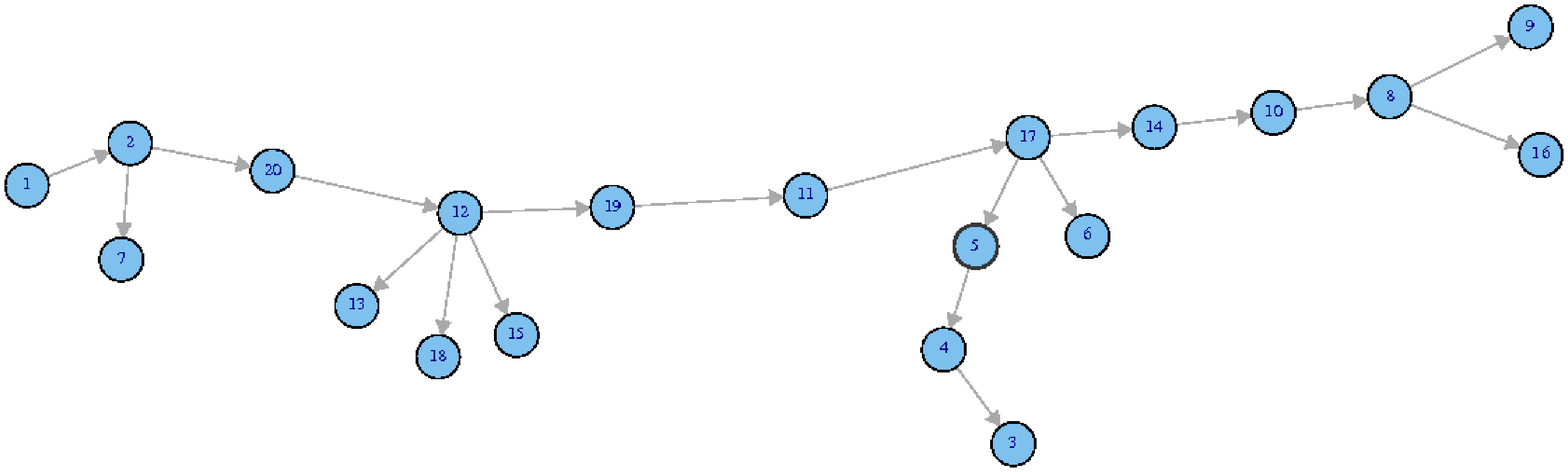}}
\fbox{(b)\includegraphics[scale=0.4]{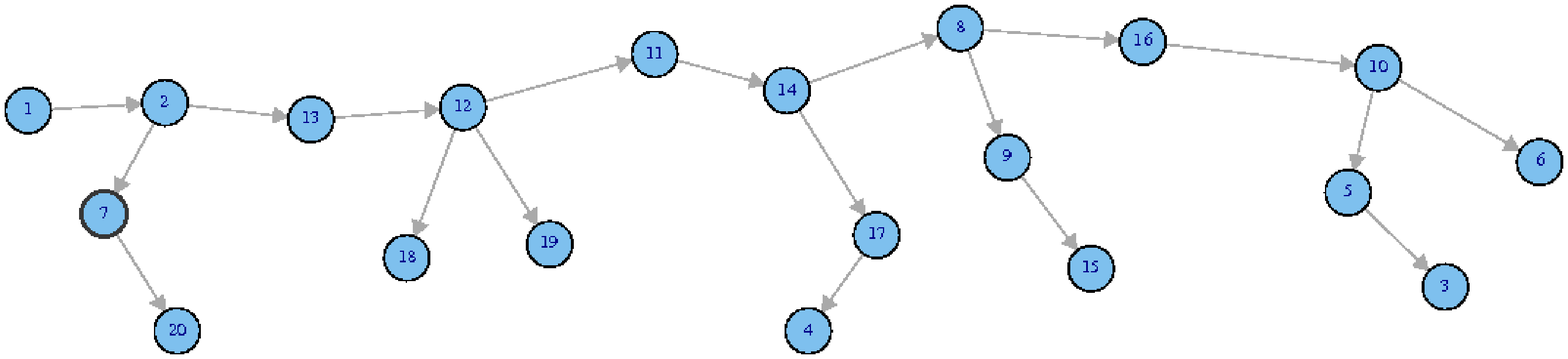}}
\caption{The MST using Prim algorithm. (a) Before the crisis (b) During the crisis. There is a strong tendency for financial indices to organize by geographical location.}
\label{mstbeforeduring}
\end{figure}

We construct the network of 20 financial indices (before and during 2008 crisis) by using the metric
distances \cite{mantegna00} $d_{ij}=\sqrt{2(1-C_{ij})}$ forming a N x N distance matrix D whose elements varies between
0 and 2. Here $C_{ij}$ is the correlation between indices i and j whose elements varies from -1 to 1 thus small
values of $d_{ij}$ imply high correlations among indices. The number of possible nodal connections of financial indices is
large, $N(N-1)/2$. The MST can reduce this complexity by showing only the $N-1$ most important non-redundant connections
in a graphical manner. We use the Prim Algorithm \cite{prim1957} for drawing MST.
Prim algorithm is an algorithm in graph theory that finds a minimum spanning tree for a connected graph i.e. it
finds a subset of the edges that forms a tree that includes every vertex, where the total weight of all the edges in the
tree is minimized. If the graph is not connected, then it will only find a MST for one of the connected components. The MST
shows the presence of clusters of nodes (indices) which are quite homogeneous and it also displays a structure in
subclusters where nodes are indices belonging to the same subsector. We find that there is a strong tendency for financial
indices to organize by geographical location that can be seen in Fig.~\ref{mstbeforeduring} (a) and (b). Before the crisis the
structure of MST is more star like whereas during the crises it changes to be more chain like. Using MST, we find that there is a strong tendency for financial indices to organize by geographical location.

\section{Multifractal Detrended Fluctuation Analysis}
\begin{figure}
\centering
\includegraphics[scale=0.4]{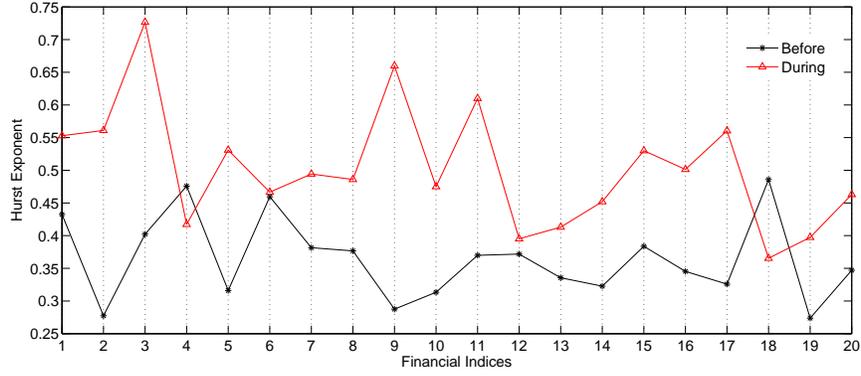}
\caption{Hurst Exponents for 20 financial indices. Hurst exponents increases for most of the financial indices during the crisis period.}
\label{Hcrisis}
\end{figure}

\begin{figure}
\centering
\includegraphics[scale=0.5]{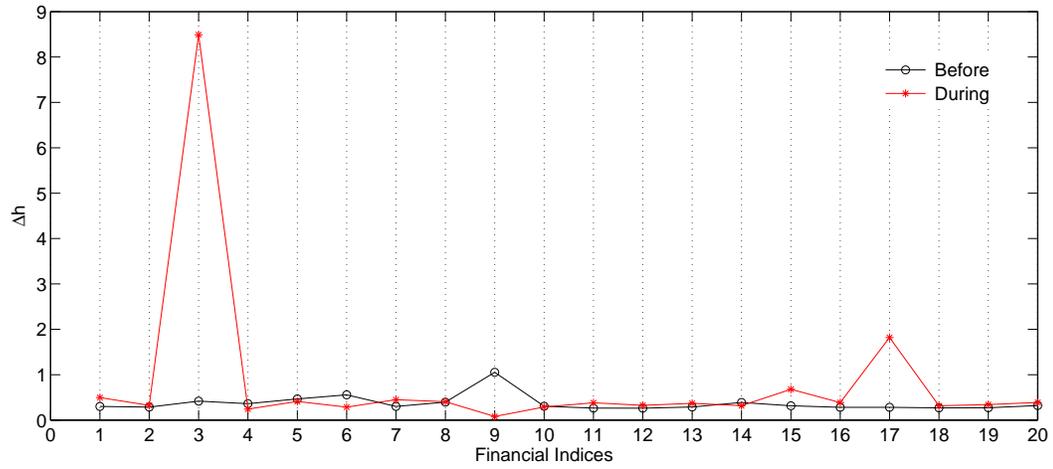}
\caption{Multifractal degree $(\Delta h)$ before and during the financial crisis for 20 financial indices. A large
variation in the value of $\Delta h$ is observed in case of Egypt, Malaysia, Taiwan, Israel and Singapore during the
crisis period.}
\label{deltah}
\end{figure}

\begin{figure}
\centering
\includegraphics[scale=0.4]{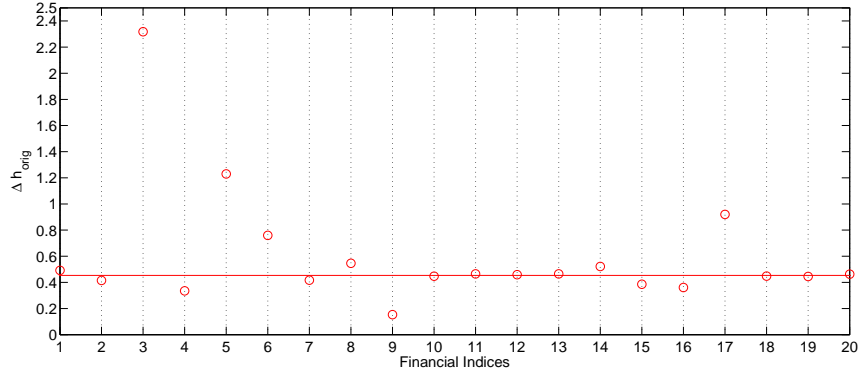}
\caption{Multifractal degree ($\Delta h$) for original return series over the full period. The line at $\Delta h=0.4516$
corresponds to indices of (Americas and Europe) which are lying in the same range of degree of multifractality. India,
South Korea, Hong Kong are found to be near the degrees of multifractality of indices corresponding to Americas and Europe. A
large variation in degrees of multifractality in Egypt, Indonesia, Malaysia, Taiwan and Singapore may be a reason that when
we increase the threshold in financial network these countries start getting disconnected at low threshold from the correlation network of financial indices.}
\label{deltahfullperiodcomparision}
\end{figure}

We define the normalized logarithmic returns as $ g_{t} = \frac{log S(t+1) - log S(t)}{\sigma} $ of
length N, where $ S(t) $ denotes the daily closing prices of the index and $ \sigma $ is the standard deviation of
logarithmic returns. In order to study the multifractal properties of 20 financial time series, we use the MF-DFA
method \cite{kantelhardt02} which consists of five steps.\\
Step 1: Calculate the "profile", $ Y(i)\equiv\sum_{k=1}^i[g_{k}-<g>], i=1,...,N,$ where N is the length of series
and $ <g> $ is the mean of $ g_{t} $.\\
 Step 2: Divide the profile $ Y(i) $ into $ N_{s}\equiv{int(N/s)} $ non-overlapping segments of equal length $ s $. Since
the length of the series is often not a multiple of the considered time scale s, a short part of the series remains, the
same procedure is repeated starting from the opposite end. Thereby, $ 2N_{s} $ segments are obtained altogether.\\
 Step 3: Calculate the local trend for each of the $2N_{s}$ segments
by a least-square fit of the time-series. Then determine the variance
$ F^{2}(s,\nu)  \equiv  \frac{1}{s}\sum_{i=1}^s  \lbrace Y[(\nu-1)s+i]-y_\nu (i) \rbrace^2 $ for
each segment $\nu$, $\nu=1,...,N_{s}$
and $F^{2}(s,\nu)\equiv \frac{1}{s}\sum_{i=1}^s \lbrace Y[N-(\nu-N_{s})s+i]-y_{\nu}(i)\rbrace^{2}$
for $ \nu=N_{s}+1,...,2N_{s}$. Here, $ y_{\nu}(i) $ is the fitting polynomial in segment $ \nu $.\\
Step 4: Average over all segments to obtain the $ q^{th}$ order fluctuation
function,$F_{q}(s)\equiv \lbrace \frac{1}{2N_{s}} \sum_{\nu=1}^{2N_{s}}[F^{2}(s,\nu)]^{q/2}\rbrace^{1/q}$ here, the
variable $ q $ can take any real value except zero \cite{kantelhardt02}.\\
Step 5: Determine the scaling behavior of the fluctuation  functions by analyzing log-log plot of $F_{q}(s)$ versus $s$ for each value of $q$. If the time series $g_{t}$ are long-range power-law correlated, $F_{q}(s)$ increases for large value of s, as a power-law $F_{q}(s)  \sim s^{h(q)}$. By construction, $F_{q}(s)$ is defined for $s\geq m+2$. The family of scaling exponents h(q) can be obtained by observing the slope of the log-log plot of $F_{q}(s)$ versus s. h(q) is the generalization of the Hurst exponent $H(\equiv h(2))$. The monofractal time series are characterized by a single exponent over all time scales i.e. $ h(q) $ is independent of $ q $, whereas for  multifractal time series, $ h(q) $ varies with $ q $.  Obviously, richer multifractality corresponds to higher variability of $h(q)$. Then, the multifractality degree can be quantified by $\Delta h=h(q_{min})-h(q_{max})$. As the large fluctuations are characterized by smaller scaling exponent $h(q)$ than small fluctuations therefore $h(q)$ for $q<0$ are larger than those for $q>0$ and $\Delta h$ is positively defined.
We calculate Hurst exponents for financial indices before and during the crisis. Fig.~\ref{Hcrisis} show increase in the value of Hurst exponents for most of the financial indices during the crisis as compared to period before the crisis. We also find the multifractal degree ($\Delta h$) of financial
indices before and during the crisis and results are compared in Fig.~\ref{deltah}. Here, we see that there is no
significant variation in the multifractal degree except the indices of Egypt, Malaysia, Taiwan, Israel, and Singapore.
We also study multifractal properties of 20 financial indices for entire period i.e. July 2, 1997 to June 1, 2009 (due to large number of figures these results are not shown instead we compared multifractal results in the Table.1). To find the origin
of multifractality in financial time series, we have shuffled these series. In the shuffling procedure the data are put
in the random order. So, all temporal correlations are destroyed without effecting the probability density function. In
order to quantify the influence of the fat-tail distribution, we generate the surrogate time series from original series
by using the Schreiber method \cite{schreiber96}. This algorithm for generating the surrogate data is based on a simple
iteration scheme called Iterated Amplitude-Adjusted Fourier Transform (IAAFT), which is an improved version of the phase
randomization algorithm \cite{theiler92}. In Table 1, we compare the multifractal degrees for original, shuffled, and surrogated
time series respectively. We find that there is a contribution of long-range correlation as well as broad probability density
function in multifractality of all financial indices except the Taiwan index where the multifractal degree for shuffled and surrogate series are weaker than those of original series. In Fig. ~\ref{deltahfullperiodcomparision}, we find that financial indices corresponds to Americas and Europe almost lie in the same range of degree of multifractality as compared to other
indices. India, South Korea, and Hong Kong are found to be near the degree of multifractality of indices corresponds to
Americas and Europe. A large variation in degree of multifractality in Egypt, Indonesia, Malaysia, Taiwan and Singapore
may be a reason that when we increase the threshold in financial network these countries start getting disconnected at low threshold from the correlation network of financial indices. We compare multifractal results of financial indices with the Binomial Multifractal Model (BMFM) \cite{feder88,barbasibmfm91,kantelhardt02}. A binomial multifractal series of $N=2^{n_{max}}$ numbers $k$ with $k=1,...,N$ is defined by $x_{k}=a^{n(k-1)}(1-a)^{n_{max}-n(k-1)}$, where $0.5<a<1$ is a parameter and $n(k)$ is the number of digits equal to $1$ in the binary representation of the index k, for example, n(19)=3 because 19 corresponds to
binary 10011. We generate binomial multifractal series with $n_{max}=12$ and different values of $a$, then compare their
multifractal results with financial indices. We find that at $a=0.4125$, the BMFM fit well to indices corresponding to the
America, Europe and Australia i.e. these indices exhibit a common multifractal behavior as compared with other indices. Other
indices also fit with BMFM for different values of parameter $a$ as follows: India, Japan, and Israel at $0.45$, South Korea
and Hong Kong at $0.6$, Indonesia at $0.725$, Malaysia at $0.65$, Singapore at $0.675$ and Egypt at $0.85$.

\section{Conclusion}
We study results obtained before and during the financial crisis of 2008 by using three methods: (i) RMT (ii) Network (Threshold and MST)  and (iii) MF-DFA. We further apply
the RMT and MF-DFA method to these indices over the entire period and study their correlation and multifractal properties. A sliding window of 25 days is used to investigate the fluctuations in financial indices. The empirical results verify the
validity of the measures, and this has led to a better understanding of complex financial markets. We analyze the eigenvalue dynamics of correlation matrix  of 20 financial indices using a sliding window of 25 days. We find that largest eigenvalues deviate significantly from the upper bound $\lambda^{rand}_{max}$ which shows a strong correlation between financial indices.
We find that the largest eigenvalue represent the collective information about the correlation between different indices and its
trend indicates the market conditions. We also perform the eigenvector analysis corresponding to the first, second and third
largest eigenvalue before and during the crisis. It is confirmed that eigenvectors corresponding to second largest eigenvalue
gives useful information about the sector formation in the global financial indices. We compare eigenvectors corresponding to
second largest eigenvalue before and during the financial crisis. Countries corresponding to financial indices above eigenvector
threshold 0.15 that are contributing more are as follows: Argentina, Brazil, Mexico, France, Germany, Switzerland, UK,
US (before the crisis) and Indonesia, Malaysia, South Korea, Taiwan, Australia, Hong Kong, Japan,
Singapore (during the crisis). We find that these sectors are formed on the basis of the geographical location.  However, eigenvectors corresponding to third largest eigenvalue does not give much information as third largest eigenvalue is near the random matrix bound.\\
We study properties of the correlation networks of 20 financial indices by using the threshold and hierarchical (MST) method respectively. We analyze the effect of financial crisis of 2008 on the correlation network of global financial indices. By constructing networks for different values of threshold $\theta$ in the range 0 to 0.9, we find that at threshold $\theta=0.2$ the network is fully connected. At threshold $\theta=0.6$, we find that the Americas, Europe and Asia/pecific form different clusters before the crisis but during the crisis Americas and Europe are strongly linked. If we further increase the threshold $\theta$ up to 0.9 we find that European countries France, Germany and UK consistently constitute the most tightly linked markets before and during the crisis. We also study the topological properties (mean degree, clustering coefficients, connected components, and clique) of correlation network before and during the crisis. Before the crisis the
structure of MST is more star like whereas during the crises it changes to be more chain like. Using MST, we find that there is a strong tendency for financial indices to organize by geographical location. We study multifractal properties of 20 financial indices. A change in the value of Hurst exponent before and during the crisis is observed for financial indices. We compare the multifractal degrees for original, shuffled, and surrogated time series respectively and find that there is a contribution of
long-range correlation as well as broad probability function in the multifractality of financial indices except the index of Taiwan as multifractal degree for shuffled and surrogate series are weaker than those of original series. We find that financial indices corresponds to Americas and Europe almost lie in the same range of degrees of multifractality. India, South Korea, Hong Kong are found to be near the degrees of multifractality of indices corresponds to Americas and Europe. A large variation in degrees of multifractality in Egypt, Indonesia, Malaysia, Taiwan and Singapore may be a reason
that when we increase the threshold in financial network these countries start getting disconnected at low threshold from the correlation network of financial indices. We fit the Binomial multifractal model to financial indices.


\begin{acknowledgments}
We would like to thank Prof. Sanjay Jain for encouragement and discussions. We acknowledge the University Faculty R\&D Grant for financial support.
\end{acknowledgments}

\clearpage
\begin{figure}
\fbox{(a)\includegraphics[scale=0.2]{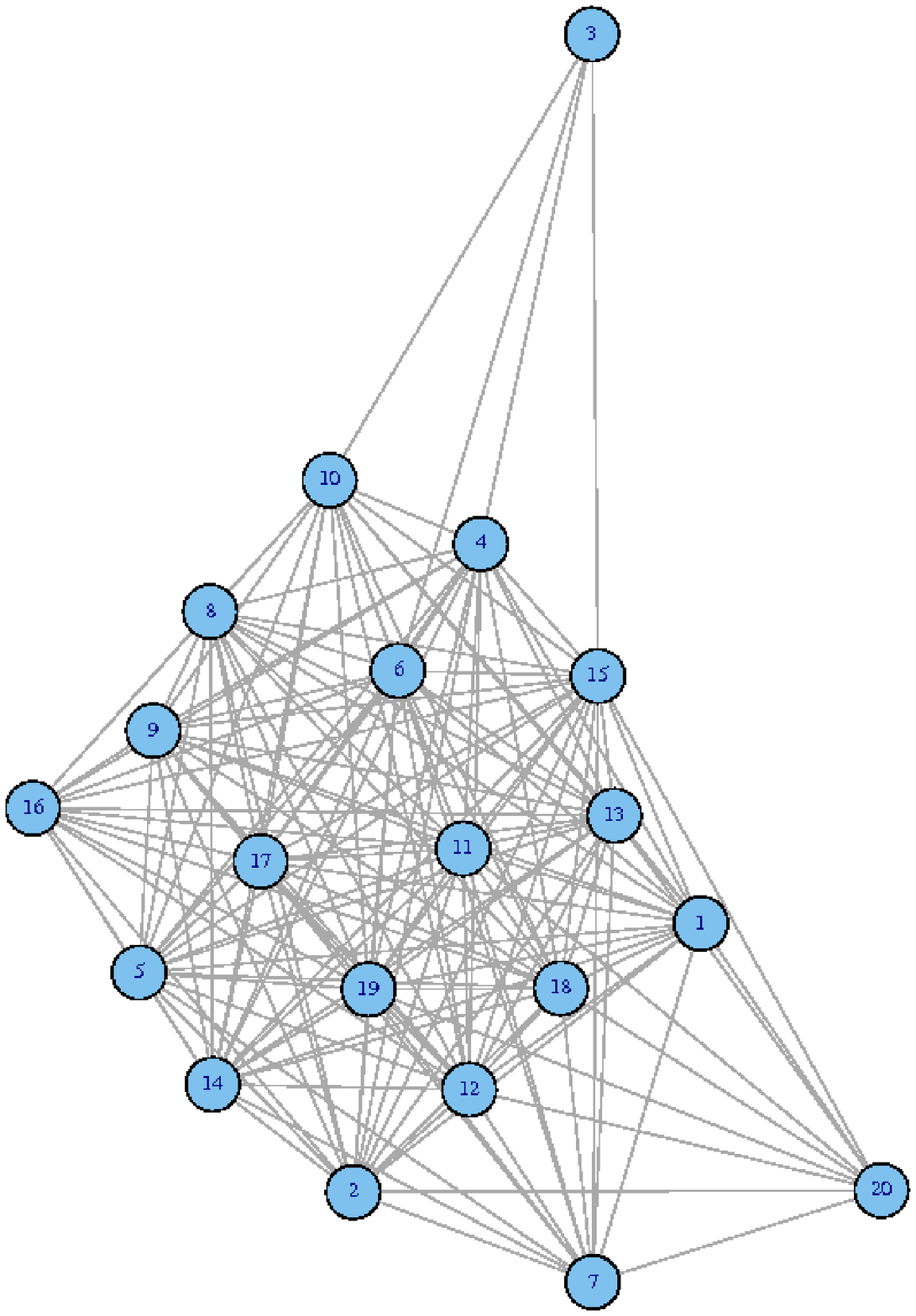}}
\fbox{(b)\includegraphics[scale=0.2]{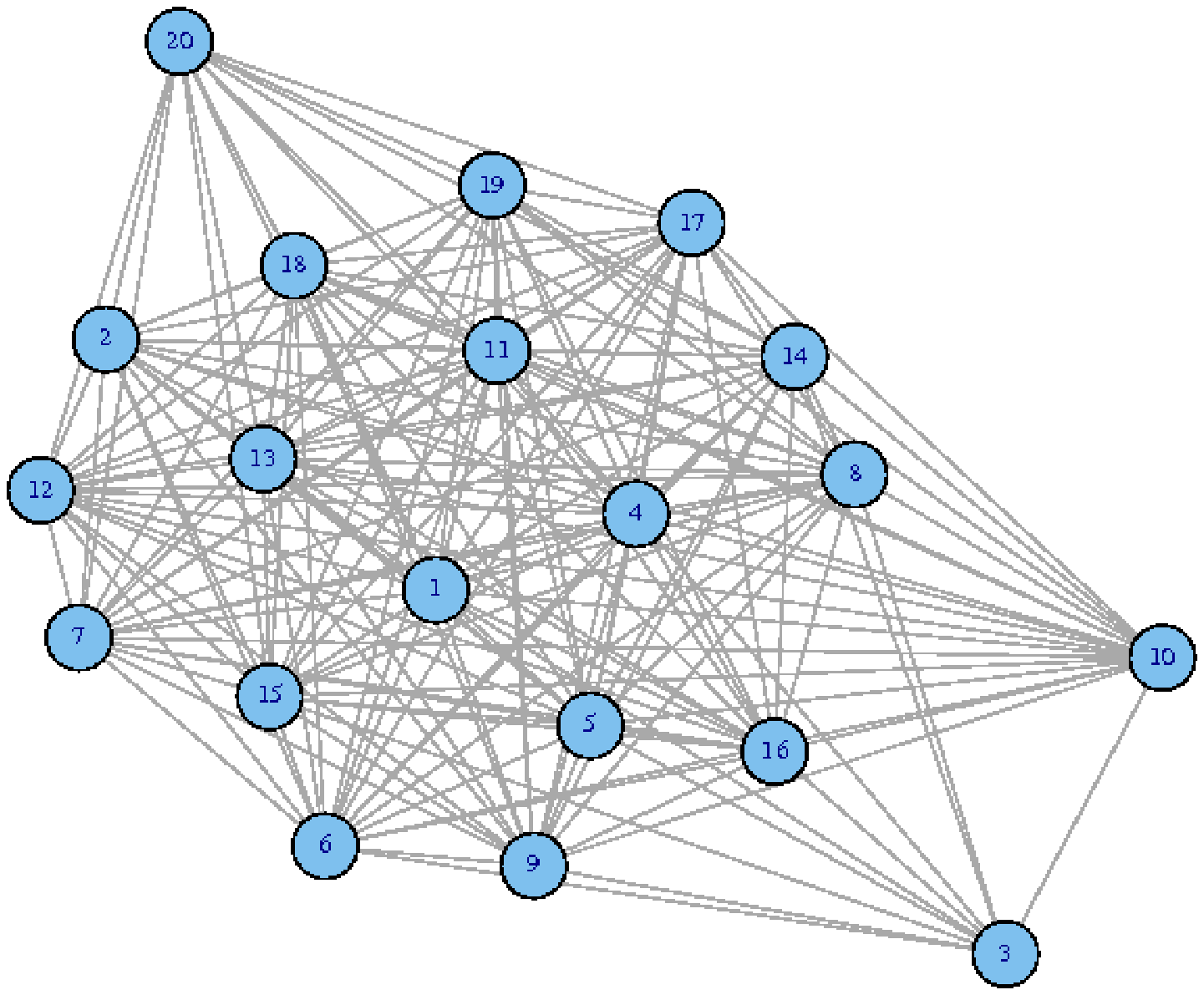}}
\fbox{(c)\includegraphics[scale=0.2]{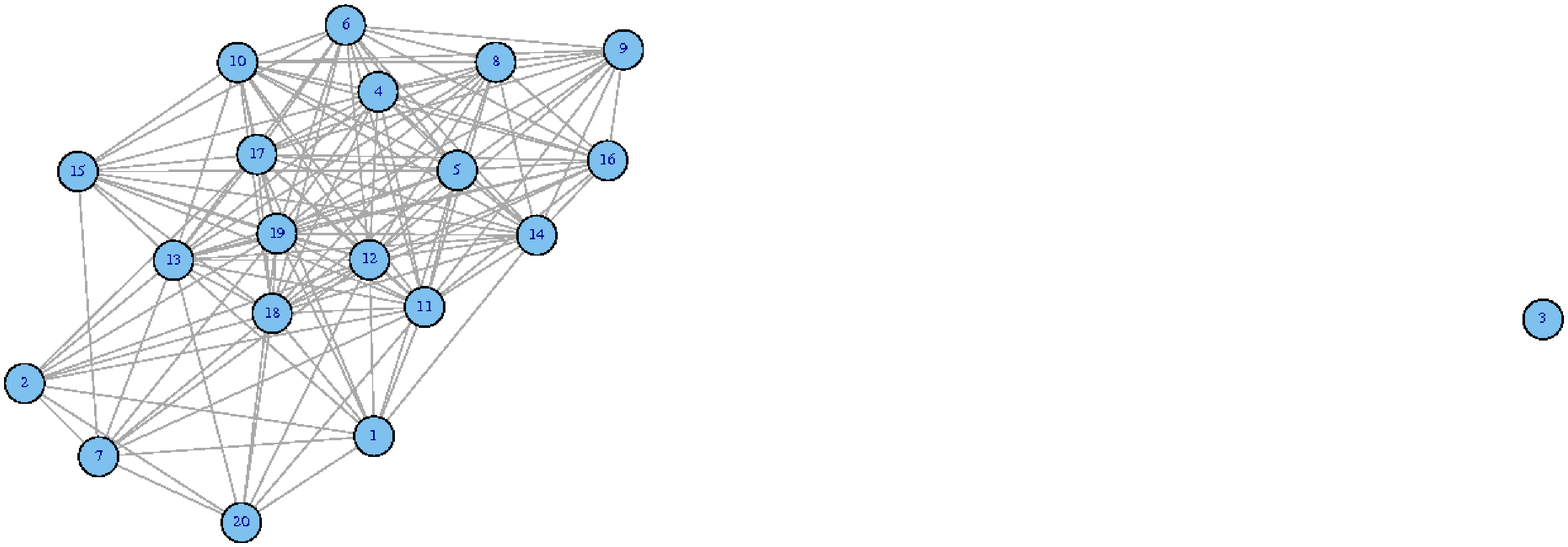}}
\fbox{(d)\includegraphics[scale=0.2]{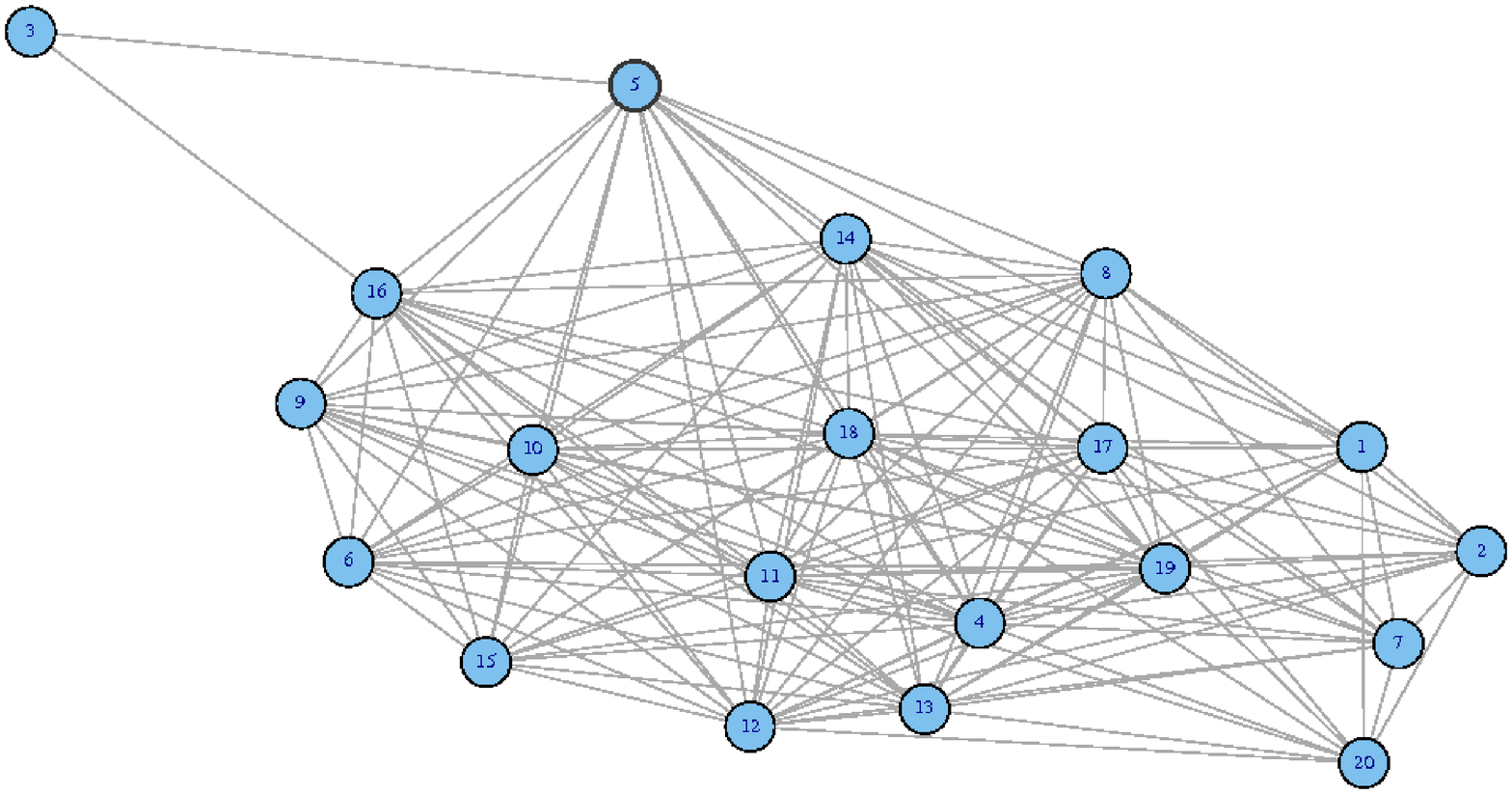}}
\fbox{(e)\includegraphics[scale=0.2]{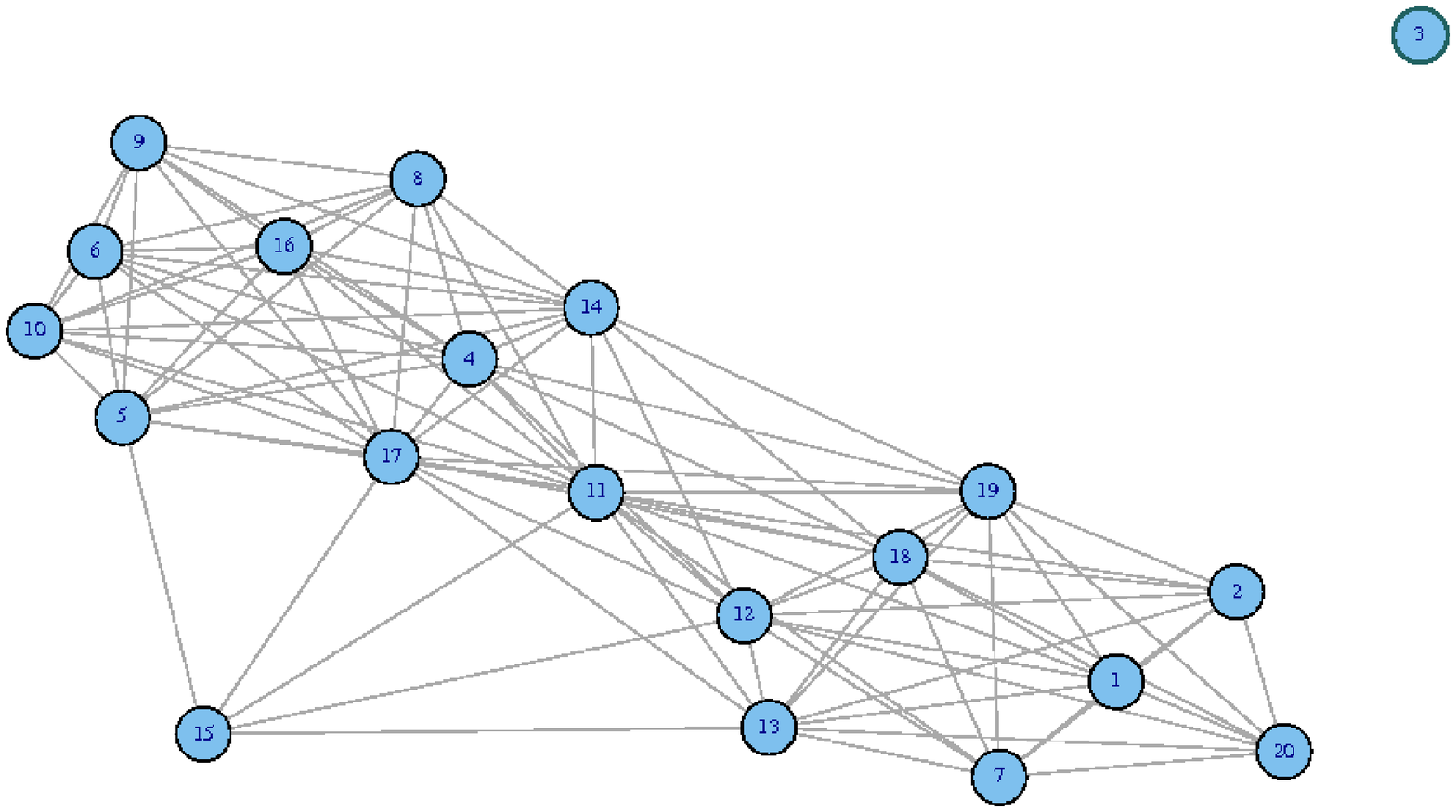}}
\fbox{(f)\includegraphics[scale=0.2]{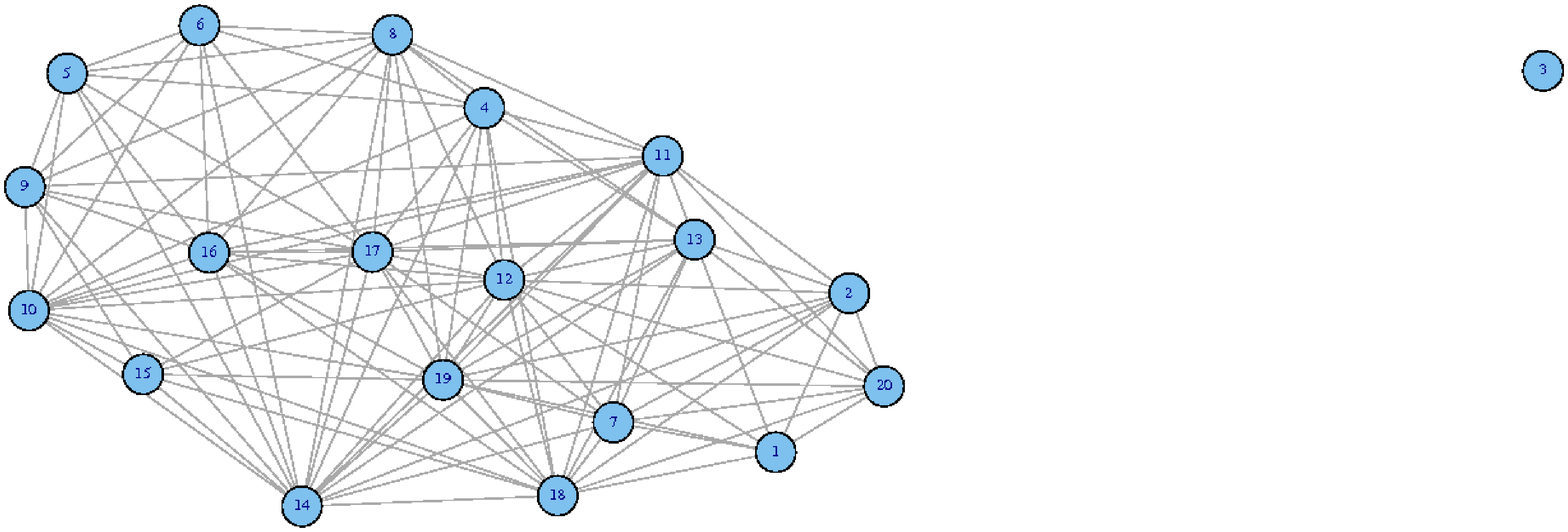}}
\fbox{(g)\includegraphics[scale=0.2]{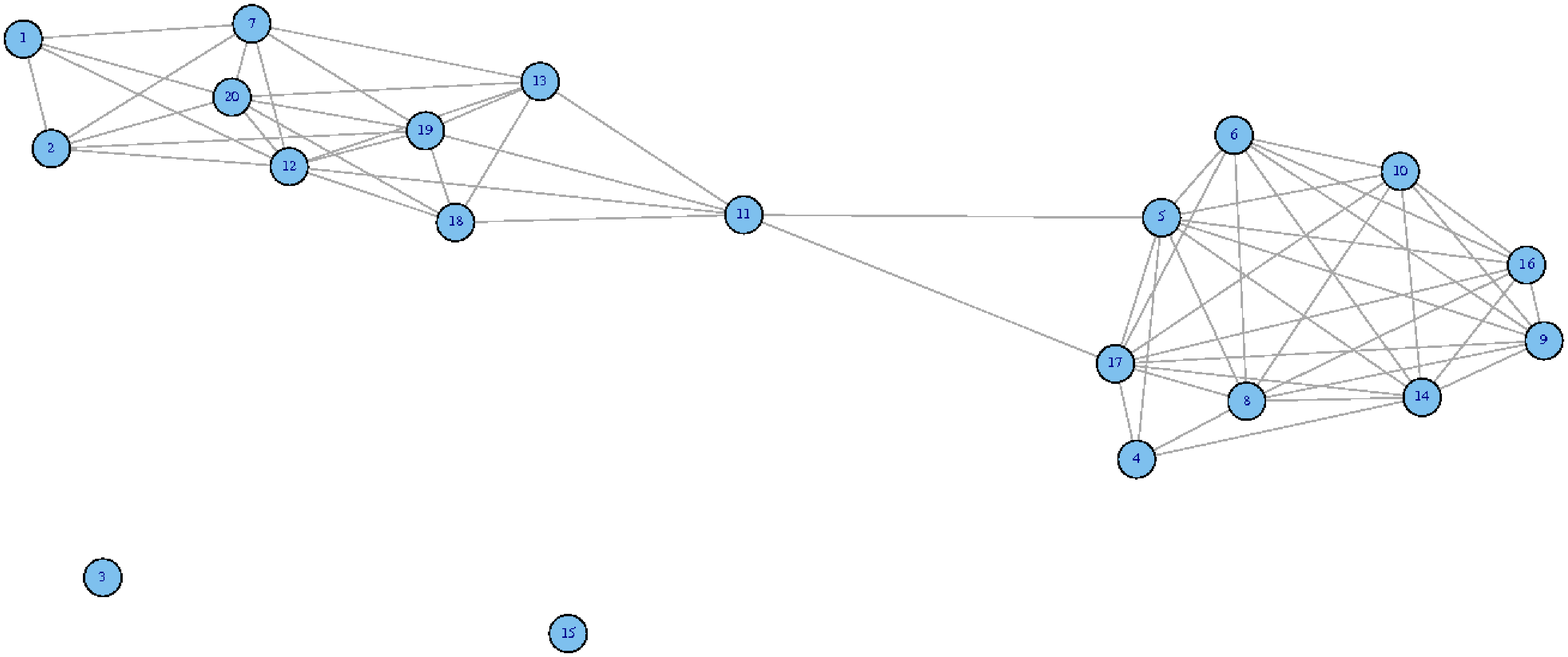}}
\fbox{(h)\includegraphics[scale=0.2]{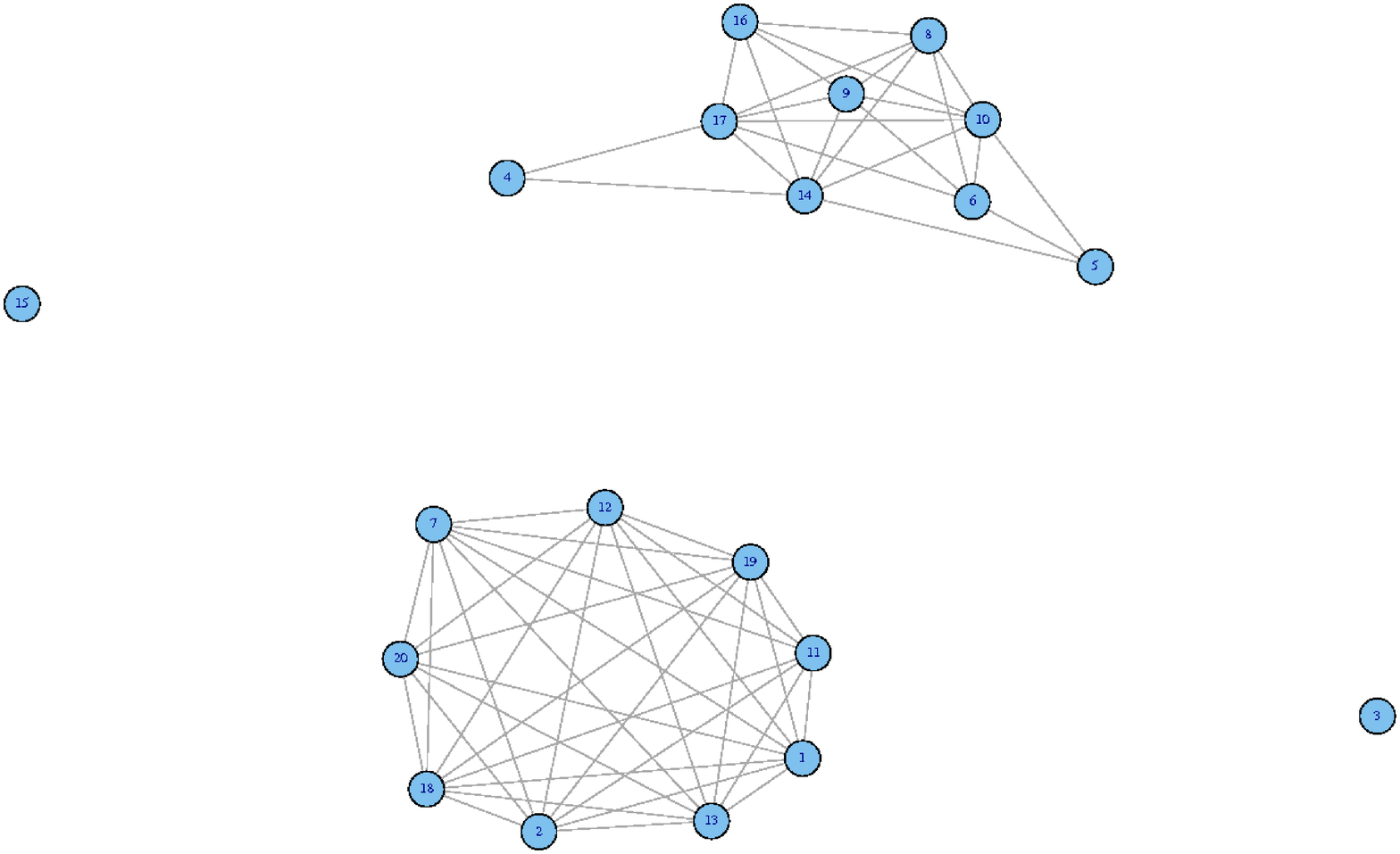}}
\caption{The financial network of 20 indices at different threshold before and during
the crisis: (a) $\theta=0.2$ (before) (b) $\theta=0.2$ (during) (c) $\theta=0.3$ (before) (d) $\theta=0.3$ (during
(e) $\theta=0.4$ (before) (f) $\theta=0.4$ (during) (g) $\theta=0.5$ (before) (h) $\theta=0.5$ (during).}
\label{thetap4to5}
\end{figure}
\newpage
\newcommand{\btable}[3]{
\begin{table}[t]
\begin{center}
\caption{#2\label{#3}}
\begin{tabular}{#1}}
\newcommand{\etable}{\end{tabular}
\end{center}
\end{table}}
\newcommand{\mc}[2]{\multicolumn{#1}{c}{#2}}
\btable{cccccccccc}{Comparison of Degrees of multifractality and volatility using the MF-DFA of 20 financial indices for full period. Financial indices corresponding to the Americas (Argentina, Brazil, Mexico, and US) behaves almost in the same range degree of multifractality. Similarly indices corresponding to Europe (Austria, France, Germany, Switzerland, UK) belongs to same range of degree of multifractality. However for Asia/Pacific (Australia, Hong Kong, India, Indonesia, Malaysia, Japan, Singapore, South Korea, Taiwan) and Africa/Middle East (Egypt, Israel), we do not find similar properties due to large variation in their multifractal degrees and volatilities.}{results}\\ \hline
S.No.&    Country & \mc{2}{$\Delta h_{orig}$} &  \mc{2}{$\Delta h_{shuf}$} &\mc{2}{$\Delta h_{sur}$} &\mc{2}{volatility} \\ \hline
 1   & Argentina   & \mc{2}{0.492 }  & \mc{2}{0.266 }  &\mc{2}{0.306 }   &\mc{2}{0.0153 }\\
 2   & Brazil      & \mc{2}{0.415 }  & \mc{2}{0.240 }  &\mc{2}{0.223 }   &\mc{2}{0.0161 }\\
 3   & Egypt       & \mc{2}{2.318 }  & \mc{2}{0.377 }  &\mc{2}{0.494 }   &\mc{2}{0.0055 }\\
 4   & India       & \mc{2}{0.335 }  & \mc{2}{0.236 }  &\mc{2}{0.241 }   &\mc{2}{0.0123}\\
 5   & Indonesia   & \mc{2}{1.230 }  & \mc{2}{0.319 }  &\mc{2}{0.321 }   &\mc{2}{0.0119} \\
 6   & Malaysia    & \mc{2}{0.760 }  & \mc{2}{0.335 }  &\mc{2}{0.358 }   &\mc{2}{0.0090 }\\
 7   & Mexico      & \mc{2}{0.417 }  & \mc{2}{0.263 }  &\mc{2}{0.312 }   &\mc{2}{0.0115 }\\
 8   & South Korea & \mc{2}{0.546 }  & \mc{2}{0.310 }  &\mc{2}{0.219 }   &\mc{2}{0.0145 }\\
 9   & Taiwan      & \mc{2}{0.153 }  & \mc{2}{0.267 }  &\mc{2}{0.237 }   &\mc{2}{0.0115 }\\
 10  & Australia   & \mc{2}{ 0.447}  & \mc{2}{0.230 }  &\mc{2}{0.267 }   &\mc{2}{0.0067 }\\
 11  & Austria     & \mc{2}{ 0.465}  & \mc{2}{0.291 }  &\mc{2}{0.326 }   &\mc{2}{0.0093 }\\
 12  & France      & \mc{2}{0.459 }  & \mc{2}{0.238 }  &\mc{2}{0.252 }   &\mc{2}{0.0108 }\\
 13  & Germany     & \mc{2}{ 0.465}  & \mc{2}{0.256}   &\mc{2}{0.255 }   &\mc{2}{0.0117 }\\
 14  & Hong Kong   & \mc{2}{0.522 }  & \mc{2}{0.254 }  &\mc{2}{0.254 }   &\mc{2}{ 0.0122}\\
 15  & Israel      & \mc{2}{0.386 }  & \mc{2}{0.327 }  &\mc{2}{0.324 }   &\mc{2}{0.0090 }\\
 16  & Japan       & \mc{2}{0.361 }  & \mc{2}{0.223 }  &\mc{2}{0.217 }   &\mc{2}{0.0111 }\\
 17  & Singapore   & \mc{2}{0.920 }  & \mc{2}{0.217 }  &\mc{2}{0.267 }   &\mc{2}{0.0101 } \\
 18  & Switzerland & \mc{2}{0.448 }  & \mc{2}{0.283 }  &\mc{2}{0.216 }   &\mc{2}{0.0093}\\
 19  & UK          & \mc{2}{0.445 }  & \mc{2}{0.243 }  &\mc{2}{0.246 }   &\mc{2}{0.0090 } \\
 20  & US          & \mc{2}{0.463 }  & \mc{2}{0.247 }  &\mc{2}{0.221 }   &\mc{2}{0.0091 }\\   \hline
\etable
\end{document}